\documentclass[12pt,preprint]{emulateapj}
\def\dave{Dav{\'e}}

\begin{document}
\title{$z\sim7-10$ Galaxies in the HUDF and GOODS fields, and their UV
Luminosity Functions}
\author{Rychard J. Bouwens$^{2}$, Garth D. Illingworth$^{2}$, 
Marijn Franx$^{3}$, Holland Ford$^{4}$
}

\affil{1 Based on observations made with the NASA/ESA Hubble Space
Telescope, which is operated by the Association of Universities for
Research in Astronomy, Inc., under NASA contract NAS 5-26555. These
observations are associated with programs \#7235, 7817, 9425, 9575, 9797, 9803, 9978, 9979, 10189, 10339, 10340, 10403, 10530, 10632, 10872, and 11082.
Observations have been carried out using the Very Large Telescope at
the ESO Paranal Observatory under Program ID(s): LP168.A-0485.  Based
in part on data collected at the Subaru Telescope and obtained from
SMOKA, which is operated by the Astronomy Data Center, National
Astronomical Observatory of Japan.}
\affil{2 Astronomy Department, University of California, Santa Cruz,
CA 95064}
\affil{3 Leiden Observatory, Postbus 9513, 2300 RA Leiden, Netherlands} 
\affil{4 Department of Physics and Astronomy, Johns 
Hopkins University, 3400 North Charles Street, Baltimore, MD 21218}

\begin{abstract}
We use all available deep optical ACS and near-IR data over both the
HUDF and the two GOODS fields to search for star-forming galaxies at
$z\gtrsim7$ and constrain the $UV$ LF within the first 700 Myrs.  Our
data set includes $\sim23$ arcmin$^2$ of deep NICMOS $J+H$ data and
$\sim248$ arcmin$^2$ of ground-based (ISAAC+MOIRCS) data, coincident
with ACS optical data of greater or equal depths.  In total, we find 8
$<$$z$$>$$\sim$7.3 $z$-dropouts in our search fields, but no $z\sim9$
$J$-dropout candidates.  A careful consideration of a wide variety of
different contaminants suggest an overall contamination level of just
$\sim12$\% for our $z$-dropout selection.  After performing detailed
simulations to accurately estimate the selection volumes, we derive
constraints on the UV LFs at $z\sim7$ and $z\sim9$.  For a faint-end
slope $\alpha=-1.74$, our most likely values for $M_{UV}^*$ and
$\phi^*$ at $z\sim7$ are $-19.8\pm0.4$ mag and
$1.1_{-0.7}^{+1.7}\times10^{-3}$ Mpc$^{-3}$, respectively.  Our search
results for $z\sim9$ $J$-dropouts set a $1\sigma$ lower limit on
$M_{UV}^*$ of $-19.6$ mag assuming that $\phi^*$ and $\alpha$ are the
same as their values at slightly later times.  This lower limit on
$M_{UV}^*$ is 1.4 mag fainter than our best-fit value at $z\sim4$,
suggesting that the $UV$ LF has undergone substantial evolution over
this time period.  No evolution is ruled out at 99\% confidence from
$z\sim7$ to $z\sim6$ and at 80\% confidence from $z\sim9$ to $z\sim7$.
The inferred brightening in $M_{UV}^*$ with redshift (i.e., $M_{UV}
^{*} = (-21.02\pm0.09) + (0.36\pm0.08) (z - 3.8)$) matches the
evolution expected in the halo mass function, if the mass-to-light
ratio of halos evolves as $\sim(1+z)^{-1}$.  Finally, we consider the
shape of the $UV$ LF at $z\gtrsim5$ and discuss the implications of
the Schechter-like form of the observed LFs, particularly the
unexpected abrupt cut-off at the bright end.
\end{abstract}

\section{Introduction}

One of the most important goals of observational cosmology has been to
identify galaxies at the earliest times and to characterize their
evolution.  Over the past few years, substantial progress has been
made on this front, particularly at high redshifts ($z\sim4-6$).  Much
of this progress has been due to the significant quantity of deep
multi-wavelength data made possible by instruments like the Hubble
Space Telescope (HST) Advanced Camera for Surveys (Ford et al.\ 2003)
or Subaru's Suprime-Cam (Miyazaki et al. 2002).  As a result, current
selections (e.g., Yoshida et al.\ 2006; Beckwith et al.\ 2006; Bouwens
et al.\ 2007) now number in the tens of thousands of galaxies and
reach to luminosities as faint as $-16$ AB mag ($\sim$0.01 $L_{z=3}
^{*}$).  This has enabled astronomers to measure a wide-range of
properties in high-redshift galaxies and examine their evolution
across cosmic time.

One of the key observables considered in previous studies of
high-redshift galaxies is the luminosity function in the rest-frame
$UV$.  The $UV$ LF is of significant interest not only because of the
quantitative constraints it imposes on early galaxy formation, but
also because of the information it provides on the star formation rate
and $UV$ output of galaxies at the earliest times.  This latter
information is important for investigating the role galaxies may have
played in the reionization of the universe (e.g., Stanway et al.\
2003; Giavalisco et al.\ 2004b; Yan \& Windhorst 2004; Bouwens et al.\
2006).

In principle, the large statistical samples of star-forming galaxies
at $z\sim4-6$ just described allow us to determine the $UV$ LFs and
their evolution in great detail out to $z\sim6$.  Unfortunately,
despite the large sample sizes and wide luminosity ranges of current
selections, there continues to be some uncertainty regarding how the
$UV$ LF evolves at very high redshift (i.e., from $z\gtrsim6$ to
$z\sim3$).  Some groups (Shimasaku et al.\ 2005; Bouwens et al.\ 2006;
Yoshida et al.\ 2006; Ouchi et al.\ 2004a; Bouwens et al.\ 2007;
Dickinson et al.\ 2004) have argued the evolution occurs primarily at
bright end of the LF, others (Iwata et al.\ 2003; Sawicki \& Thompson
2006; Iwata et al.\ 2007) have asserted it happens more at the
faint-end, and still others have suggested the evolution is
independent of luminosity (Beckwith et al.\ 2006).  The diversity of
conclusions drawn by different teams would seem to suggest that
systematics are likely playing a large role.  A careful discussion of
this issue as well as detailed comparisons with contemporary
determinations of the $UV$ LF at $z\sim4-6$ can be found in Bouwens et
al.\ (2007).

One way of gaining additional leverage in looking at the evolution of
the $UV$ LF at high redshift is to push to even higher redshift, i.e.,
$z\sim7-10$, to see what galaxies were like at these epoches.  Since
it is expected that changes in the $UV$ LF will be much more
substantial over the interval $z\sim7-10$ to $z\sim3-4$ than from
$z\sim6$ to $z\sim4$, it should be relatively easy to ascertain how
the $UV$ LF is evolving at early cosmic times.  The challenge, of
course, is that the number of star-forming galaxy candidates at
$z\sim7-10$ has been very small, of order $\sim4-5$ galaxies (Bouwens
et al.\ 2004c; Bouwens et al.\ 2005; Bouwens \& Illingworth 2006; Iye
et al.\ 2006; Labb{\'e} et al.\ 2006; Mannucci et al.\ 2007; Stark et
al.\ 2007b; Bradley et al.\ 2008).  As a result, the bulk of the
effort has been directed towards simply establishing the overall
prevalence of bright star-forming galaxies at $z\sim7-10$ (Bouwens et
al.\ 2004c; Yan \& Windhorst 2004; Bouwens et al.\ 2005; Bouwens \&
Illingworth 2006; Mannucci et al.\ 2007; Iye et al.\ 2006; Stanway et
al.\ 2008).

In the present work, we will try to take studies of the $UV$ LF at
$z\gtrsim7$ one step further and not just look at the overall
prevalence of star-forming galaxies at $z\gtrsim7$, but at how the
volume density of these galaxies appears to vary as a function of
luminosity.  In doing so, we will attempt to derive constraints on the
overall shape of the $UV$ LF at $z\gtrsim7$ and therefore evaluate
different models for how the $UV$ LF evolves with cosmic time.  In
particular, we are interested in looking at how the volume density of
the bright population evolves and how this compares to the evolution
of the faint population.  If the mean luminosity of galaxies really
does increase with time at early cosmic times, as one might expect
from hierarchical models, the volume density of bright galaxies should
evolve much more rapidly at $z\gtrsim7$ than the volume density of the
faint population.  This change has been alternatively parametrized as
a brightening of the characteristic luminosity $M_{UV}^*$ with cosmic
time (e.g., Bouwens et al.\ 2006; Yoshida et al.\ 2006; Bouwens et
al.\ 2007) or a flattening of the faint-end slope $\alpha$ (e.g., Yan
\& Windhorst 2004).

To help with this endeavor, we will make use of the previous data used
by our group in high-redshift ($z\sim7-10$) galaxy searches (e.g.,
Bouwens \& Illingworth 2006; Bouwens et al.\ 2005) and also take
advantage of a significant amount of newer data.  Among these new data
are deep ACS observations over the NICMOS parallels to the HUDF (Oesch
et al.\ 2007), a large number of moderately deep NICMOS observations
over the two GOODS fields (H. Teplitz et al.\ 2008, in prep), the deep
wide-area ISAAC observations over the CDF-South GOODS field (Mannucci
et al.\ 2007; Stanway et al.\ 2008; B. Vandame et al.\ 2008, in
preparation; J. Retzlaff et al.\ 2008, in prep), and the deep
wide-area MOIRCS observations over the HDF-North GOODS field (Kajisawa
et al.\ 2006; Ouchi et al.\ 2007).  We will also be taking advantage
of improvements in our NICMOS reduction techniques to push our
previous selections slightly fainter.  These new data (particularly
the new wide-area ground-based data) will be crucial for examining the
UV LF at $z\gtrsim7$ over a wider range in luminosity than was
possible in previous work (e.g., Bouwens et al.\ 2004c; Bouwens \&
Illingworth 2006; Mannucci et al.\ 2007; Stanway et al.\ 2008),
providing us with additional leverage to constrain the shape of the
$UV$ LF at $\gtrsim7$.  The ultra deep ACS data over the NICMOS
parallels to the HUDF will also allow us to increase both the depth
and robustness of our searches for $z\sim8-10$ galaxies.

We will organize this paper as follows.  In \S2, we will provide a
brief description of the observations we will use in this paper.  In
\S3, we will describe how we select our dropout samples from the
imaging data and how we estimate the likely contamination levels.  In
\S4, we will use these samples to derive the rest-frame $UV$ LF at
$z\sim7$ and set constraints on this LF at $z\sim9$.  Finally, in \S5,
we discuss our results, and in \S6 we will provide a summary of the
main conclusions from this work and provide a short outlook for future
progress.  Throughout this work, we will find it convenient to denote
luminosities in terms of the characteristic luminosity $L_{z=3}^{*}$
at $z\sim3$ (Steidel et al.\ 1999), i.e., $M_{UV,AB}=-21.07$.  Where
necessary, we assume $\Omega_0 = 0.3$, $\Omega_{\Lambda} = 0.7$, $H_0
= 70\,\textrm{km/s/Mpc}$.  Although these parameters are slightly
different from those determined from the WMAP five-year results
(Dunkley et al.\ 2008), they allow for convenient comparison with
other recent results expressed in a similar manner.  Our $z$ and $J$
dropout criteria select galaxies over a range in redshift (i.e.,
$z\sim6.5-8.0$ and $z\sim8-10$, respectively), with mean redshifts
$<$$z$$>$=7.3 and $<$$z$$>$=9.0, respectively (see \S4.2).  For
succintness, we will usually refer to the redshift ranges selected by
these dropout criteria as $z\sim7$ and $z\sim9$.  We will express all
magnitudes in the AB system (Oke \& Gunn 1983).

\section{Observational Data}

A summary of the observational data used to search for $z\gtrsim7$ $z$
and $J$ dropouts is provided in Table~\ref{tab:obsdata}.

\subsection{NICMOS Search Fields with Deep $zJH$ coverage}

Our deepest, wide-area constraint on the rest-frame $UV$ LF at
$z\sim7-10$ comes from the numerous HST NICMOS NIC3 fields which lie
in and around the two GOODS fields.  Our reductions of the NICMOS data
over and around the GOODS fields (including the HUDF NICMOS data) was
performed with the python package ``nicred.py'' (Magee et al.\ 2007)
and represent a slight update to the reductions we performed for the
Bouwens \& Illingworth (2006) work.  The most important change in our
reduction procedure is that we now include a correction for the
non-linear response of the NIC3 camera (de Jong et al.\ 2006).  This
has the effect of brightening the measured magnitudes in the $J_{110}$
and $H_{160}$ bands by $\sim0.15$ and $\sim0.05$ mag (relative to what
these magnitudes would be without the corrections).  Our reductions of
the optical ACS data are the same as those considered in our most
recent work on the UV LF at $z\sim4-6$ (Bouwens et al.\ 2007) and
include essentially all primary and parallel ACS data taken in the
vicinity of the two GOODS fields (e.g., Giavalisco et al.\ 2004a;
Bouwens et al.\ 2004a; Riess et al.\ 2007; Oesch et al.\ 2007).

\begin{figure*}
\epsscale{1.15}
\plotone{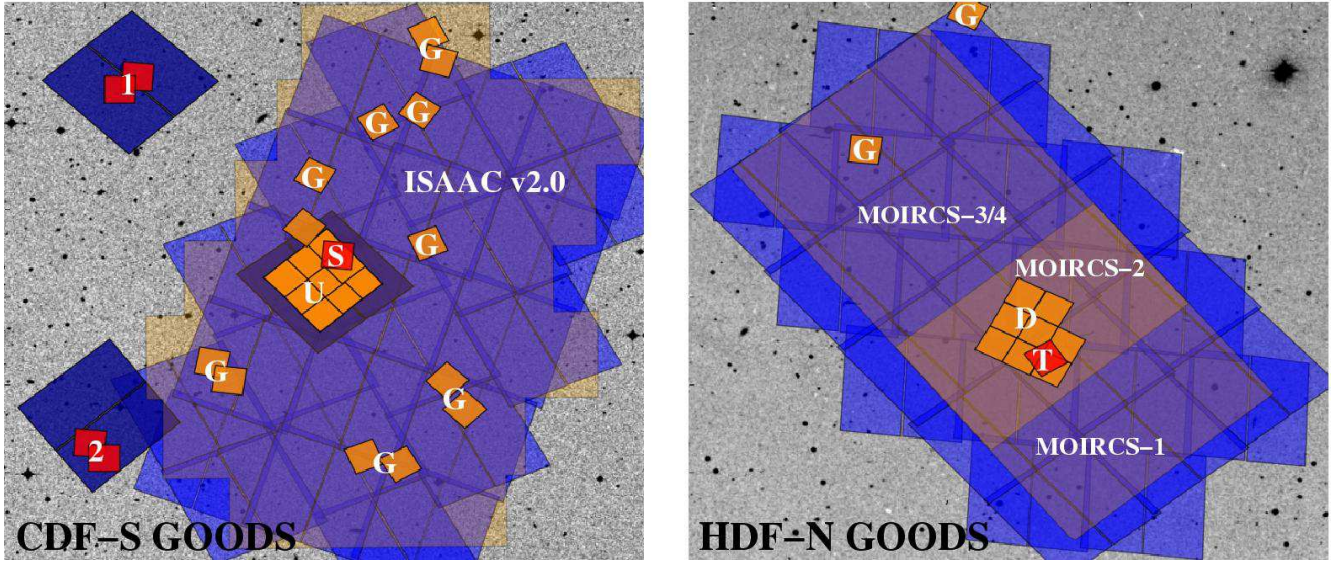}
\caption{Deep near-IR fields in and around the CDF-South
(\textit{left}) and HDF-N (\textit{right}) GOODS fields where we
conducted searches for star-forming galaxies at $z\gtrsim7$ using $z$
and $J$ dropout criteria.  Areas with deep ($\gtrsim28$ mag at
$5\sigma$) or ultra-deep ($\gtrsim29$ mag at $5\sigma$) optical
$V_{606}i_{775}z_{850}$ coverage are shaded in blue and dark blue,
respectively.  Areas with deep ($\gtrsim26.5$ mag at $5\sigma$) or
ultradeep ($\gtrsim28$ mag at $5\sigma$) NICMOS coverage are shown in
orange and red, respectively.  For clarity, we have labelled each of
these search fields with a letter, ``U'' indicating the HUDF Thompson
field (Thompson et al.\ 2005), ``S'' indicating the HUDF Stiavelli
field (Oesch et al.\ 2007), ``D'' indicating the HDF-North Dickinson
field (Dickinson 1999), ``T'' indicating the HDF-North Thompson field,
(Thompson et al.\ 1999), ``1'' indicating HUDF-NICPAR1 (Bouwens et
al.\ 2005; Oesch et al.\ 2007), ``2'' indicating HUDF-NICPAR2 (Bouwens
et al.\ 2005; Oesch et al.\ 2007), and ``G'' representing the GOODS
parallel fields (S. Malhotra et al.\ 2008, in prep; H. Teplitz et al.\
2008, in prep).  The deep, wide-area VLT ISAAC coverage of the
CDF-South GOODS field is shown as a light orange region.  The
$\sim112$ arcmin$^2$ of deep MOIRCS coverage (the 4 GTO pointings)
over the HDF-North GOODS field is shown with a similar orange shading
(the deepest MOIRCS pointing having the darkest shading).  Additional
information on our search fields can be found in
Table~\ref{tab:obsdata}.  \label{fig:obsdata}}
\end{figure*}

Our primary NICMOS data set includes a somewhat larger number of
search fields, some of significant depth, and better optical coverage
than was used in our previous search for $z\sim7$ galaxies (Bouwens \&
Illingworth 2006: see Figure 1 from the Supplementary Information).
The most notable additions are (1) one ultra deep (28.1 AB mag at
$5\sigma$) $\sim0.7$ arcmin$^2$ NIC3 pointing within the area of the
HUDF (Oesch et al.\ 2007), (2) ultra deep (28.7 AB mag at $5\sigma$)
optical ACS coverage over the second NICMOS parallel to the HUDF
(Oesch et al.\ 2007), and (3) $\sim4$ arcmin$^2$ of additional deep
(26.7 AB mag at $5\sigma$) NICMOS data within the GOODS area (Riess et
al.\ 2007; H. Teplitz et al.\ 2008, in preparation).

We have also made improvements to our reductions of the two deep
NICMOS fields taken in parallel to the HUDF (Bouwens \& Illingworth
2006; Oesch et al.\ 2007: see Figure~\ref{fig:obsdata}).  Reducing the
NICMOS data over these fields is challenging because of the small
number of unique dither positions used in the acquisition of the data
and the large number of individual frames ($\sim200$ frames in each
field).  As a result, any systematics present in characterizing the
properties of individual pixels on the NIC3 detector propagate
directly into the reduced frames, and therefore it becomes necessary
to calibrate these properties sufficiently well so that systematics do
not have a dramatic effect on the overall noise characteristics.  To
accomplish this, we generated super median images from a stack of all
the NIC3 frames taken over a field, subtracted this median image from
each NIC3 frame, and then redrizzled the individual frames together to
produce a cleaner reduction.  Before generating the super median
images, we mask out all sources identified in the $\chi^2$ image
(Szalay et al.\ 1999) created from our initial reductions of the
$J_{110}$ and $H_{160}$ band imaging data; $\sim300-400$ sources were
masked per NIC3 frame.  We elected to use the $\chi^2$ image for this
process to ensure that the same sources were masked in both the
$J_{110}$ and $H_{160}$ frames.  Therefore, the colors of sources were
not affected by this ``median stacking and subtraction'' procedure.
To obtain the best possible reduction of the data, we repeated the
above procedure several times: using each successive reduction of the
data to better mask out the faint sources in the data, median stacking
the masked frames, subtracting the medians from the individual frames,
and then drizzling the data together to generate a cleaner ``final''
reduction.

In the dropout searches we performed over the NICMOS data, we only
included data which satisfied certain minimal depth requirements in
both the optical and near-IR data to avoid the edges of the individual
reductions where the number of individual frames contributing to any
given pixel was small (i.e., where there might be concerns about some
non-Gaussian characteristics in the noise).  In our $z\sim7$
$z_{850}$-dropout search, we only included areas where the $z_{850}$,
$J_{110}$, and $H_{160}$ depths were 27.5 AB mag, 26.8 AB mag, and
26.5 AB mag, respectively, over a 0.6$''$-diameter aperture
($5\sigma$).  We demanded greater depth for our $J_{110}$-band imaging
data than we did for the $H_{160}$-band data given the importance of a
robust detection in the $J_{110}$-band to making the case for an
object being at $z\gtrsim7$.  For our $z\sim9$ $J_{110}$-dropout
search, we only included areas where the $z_{850}$, $J_{110}$, and
$H_{160}$-band depths were 27.5 AB mag, 26.8 AB mag, and 26.8 AB mag,
respectively, over a 0.6$''$-diameter aperture ($5\sigma$).  We set
more stringent requirements on the $H_{160}$-band depths for our
$J_{110}$ dropout search because of the greater need to use the
$H_{160}$-band data to establish the reality of the sources.  In
total, the area of our search fields were $\sim23$ arcmin$^2$ for our
$z_{850}$-dropout selection and $\sim21$ arcmin$^2$ for our
$J_{110}$-dropout selection.

\begin{deluxetable*}{ccccccc}
\tablecolumns{7}
\tablewidth{4.0in}
\tablecaption{Primary imaging data used for our $z$ and $J$ dropout
searches.\tablenotemark{a}\label{tab:obsdata}}
\tablehead{
\colhead{} & \colhead{} & \multicolumn{3}{c}{5$\sigma$ Depth\tablenotemark{b}} & \colhead{} \\
\colhead{Name} & \colhead{Area} & \colhead{$z_{850}$} & \colhead{$J_{110}$} & 
\colhead{$H_{160}$} & \colhead{$K_{s}$} & \colhead{Ref\tablenotemark{c}}}
\startdata
HDF-North Dickinson & 4.0 & 27.6 & 26.8 & 26.8 & 25.4 & [1,2] \\
HDF-North Thompson & 0.8 & 27.6 & 27.8 & 27.9 & 25.4 & [3,2] \\
HUDF Thompson & 5.8 & 28.8 & 27.4 & 27.2 & 25.8 & [4,5] \\
HUDF Stiavelli & 0.7 & 28.8 & 27.9 & 27.7 & 25.8 & [5,6] \\
HUDF-NICPAR1 & 1.3 & 28.4 & 28.4 & 28.2 & -- & [6,7] \\
HUDF-NICPAR2 & 1.3 & 28.4 & 28.4 & 28.2 & -- & [6,7] \\
GOODS Parallels & 9.3 & 27.3 & 26.9 & 26.7 & $\sim25$\tablenotemark{d} & [7] \\
ISAAC v2.0 & 136 & 27.3 & $\sim25.5$\tablenotemark{d} & $\sim25.0$\tablenotemark{d} & $\sim25$\tablenotemark{d} & [8,9,10] \\
MOIRCS GTO-2 & 28 & 27.3 & 25.4 & -- & 25.4 & [2] \\
MOIRCS GTO-1,3,4 & 84 & 27.3 & 24.0 & -- & 24.2 & [2] \\
\enddata
\tablenotetext{a}{The layout of these search fields is illustrated in
Figure~\ref{fig:obsdata}.}
\tablenotetext{b}{$5\sigma$ depths for ACS and NICMOS data given in
terms of a $0.6''$-diameter aperture and in terms of a
$\sim1.0''$-diameter aperture for the ground-based $K_s$-band data.
In contrast to the detection limits quoted in some of our previous
work, here our detection limits have been corrected for the nominal
light outside these apertures (assuming a point source).  The
detection limits without this correction are typically $\sim0.2$ mag
fainter.}
\tablenotetext{c}{References: [1] Dickinson 1999, [2] Kajisawa et al.\
2006, Ouchi et al.\ 2007, [3] Thompson et al. 1999, [4] Thompson et
al. (2005), [5] Labb{\'e} et al.\ (2006), [6] Oesch et al.\ 2007, [7]
Bouwens \& Illingworth (2006), Riess et al.\ (2007), H. Teplitz et
al.\ (2008, in prep) [8] B. Vandame et al.\ 2008, in prep; J. Retzlaff
et al.\ 2007, in prep, [9] Mannucci et al.\ 2007, and [10] Stanway et
al.\ (2008).}
\tablenotetext{d}{The depth of the near-IR data over the CDF-South
varies by $\sim0.2-0.4$ mag depending upon the observational
conditions in which the ISAAC data were taken.}
\end{deluxetable*}

\subsection{ISAAC Search Fields}

The deep ($J\sim25.5$ AB mag, $H,K\sim25.0$ AB mag at $5\sigma$),
wide-area ($\sim143$ arcmin$^2$) ISAAC data over the CDF-South GOODS
field (B. Vandame et al.\ 2008, in prep; J. Retzlaff et al.\ 2008, in
prep) offer us a superb opportunity to constrain the volume density of
very bright star-forming galaxies at $z\sim7-10$.  While these data
are $\gtrsim1$ mag shallower than our NICMOS search data considered
above (\S2.1), they cover $\gtrsim5$ times more area and thus are
useful for constraining the volume density of the rarer, more luminous
$UV$-bright galaxies at $z\gtrsim7$.  These data are most useful for
searching for $z_{850}$-band dropouts due to the greater depth of the
ISAAC $J$-band data.  While Mannucci et al.\ (2007) used an earlier
reduction of these data (v1.5) to search for bright $z$ dropouts over
the CDF-South GOODS field (see also Stanway et al.\ 2008), here we
repeat this selection with our procedures.  In doing so, we will be
(1) able to more naturally include the search results over these
fields with those obtained from our NICMOS fields and (2) be able to
take advantage of the additional ISAAC data that have become available
over the CDF-S since the completion of the Mannucci et al.\ (2007)
work.  The larger data set covers $\sim$10\% more area in the $J$ and
$K_s$ bands and 30\% more area in the $H$ band.  We will not be using
these images to constrain the volume density of $z\sim9$ $J$-dropouts
because of the limited depth ($\sim25$ mag at $5\sigma$) of the ISAAC
$H$ and $K_s$ band data.

We make use of the final (v2.0) reductions of the ISAAC data
(B. Vandame et al.\ 2008, in prep; J. Retzlaff et al.\ 2008, in prep)
for our searches.  These reductions include $\sim143$ arcmin$^2$ of
deep $J$ and $K_{s}$ band coverage and $\sim131$ arcmin$^2$ of deep
$H$ band coverage.  Of the area with deep J and K$_s$ coverage, only
$\sim136$ arcmin$^2$ has deep ACS GOODS coverage.  This represents 4\%
more area than considered in the $z$-dropout search by Mannucci et
al.\ (2007) using the v1.5 ISAAC reductions over the CDF-South GOODS
field.  The FWHM of the PSF varies from $\sim0.35''$ to $\sim0.7''$,
with a median value of $\sim0.51''$.  The depth of the $J$-band data
is appropriately 25.5 AB mag for a $4\sigma$ detection, while for the
$H$ and $K_s$ band data, the depth is $\sim24.9$ AB mag.  Of course,
the actual depth varies by several tenths of a magnitude across the
mosaic depending upon the precise exposure time of the ISAAC data and
the FWHM of the PSF.  As in the previous section, we used the same
reductions of the ACS GOODS data as we used in our latest paper on the
UV LF at $z\sim4$, 5, and 6 (Bouwens et al.\ 2007).

\subsection{MOIRCS Search Fields}

We can strengthen our constraints on the prevalence of the bright
star-forming galaxies at $z\gtrsim7$ by taking advantage of the very
deep wide-area data taken over the HDF-North GOODS field.  While there
have been a number of independent attempts to obtain deep near-IR
coverage over this field (e.g., the WIRC data over the HDF-North
described by Erb et al.\ 2006), perhaps the best near-IR data set
(i.e., deepest, widest area) is from the Subaru MOIRCS imaging
observations (Kajisawa et al.\ 2006; Ouchi et al.\ 2007).  This is due
to the significant investment of telescope time ($\sim17$ nights) made
by the MOIRCS GTO team in obtaining very deep $J+K_s$ images during
March, April, and May of 2006 (additional observations were made in
March and April of 2007 but those are not yet public).  There are also
2 nights of high quality $K_s$ band data available from a separate GO
proposal.  One of the main science goals for these observations has
been to search for $z\gtrsim7$ galaxies (e.g., Ouchi et al.\ 2007).

The raw images were downloaded from the Subaru-Mitaka-Okayama-Kiso
Archive (SMOKA: Baba et al.\ 2002) and then divided into a number of
discrete segments, each segment containing $<$50 images taken one
after another in a given filter.  Images from each segment were
reduced and combined into a single image using the iraf routines in
MCSRED (I. Tanaka et al.\ 2008, in prep).  These reductions were then
aligned against the ACS GOODS data and then stacked together to
produce two very deep near-IR image mosaics in the $J$ and $K_s$
bands.  These images were flux calibrated by matching sources with
those in the KPNO images of the WFPC2 HDF-North (Dickinson 1998) and
in overlap regions between adjacent MOIRCS exposures.  Since the
MOIRCS GTO team took their observations over 4 distinct pointings, the
final reductions cover $\sim112$ arcmin$^2$ in total.  The final
reductions reach a $5\sigma$ AB magnitude of $25.4$ in the $J$ band
and 25.4 AB mag in the $K_s$ band over the deepest MOIRCS pointing
(GTO-2: $\sim28$ arcmin$^2$), while the other three pointings
(GTO-1,2,4) are some $\sim$1.3 mag shallower.  The FWHM of the PSF in
our reductions was $\sim0.5''$ in the $K_s$ band and $\sim0.6''$ in
the $J$ band.  The final $J$-band image mosaic include some $\sim$56
hours of data, and the final $K_s$ band image mosaic includes some
$\sim$64 hours of Subaru data.

\section{Sample Construction}

\subsection{Catalog Construction}

Our procedures for generating PSF-matched catalogues and selecting $z$
and $J$-dropouts are already well-documented in several previous works
(e.g., Bouwens et al. 2003; Bouwens et al.\ 2005; Bouwens \&
Illingworth 2006; Bouwens et al.\ 2007).  Nonetheless, we will briefly
summarize the different steps in our processing here.  All images were
first PSF-matched to the NICMOS $H_{160}$-band data (or $J$-band data
in the case of our dropout searches over the ISAAC or MOIRCS data).
We then ran SExtractor (Bertin \& Arnouts 1996) in double-image mode
to perform the object detection and photometry.  We generated the
detection image by taking the square root of $\chi^2$ image (Szalay et
al.\ 1999) constructed from the $J_{110}$ and $H_{160}$ band images
for our $z$-dropout selection and from the $H_{160}$ band image for
our $J$-dropout selection.  Colors were measured in a small-scalable
aperture using Kron-style (1980) photometry, where the Kron factor was
1.2.  These fluxes were then corrected up to total magnitudes using
the light within a larger Kron (1980) aperture (adopting a Kron factor
of 2.5).  These latter corrections were made from the square root of
the $\chi^2$ image to improve the S/N.  Figure 5 of Coe et al.\ (2006)
provides a graphical description of a similar multi-stage procedure
for measuring colors and total magnitudes.  Aperture radii used for
total magnitude measurements ranged from $\sim0.5-0.8$ arcsec.  An
additional $\sim0.1-0.15$ mag correction was made to account for the
light outside of these apertures and on the wings of the PSF (e.g.,
Thompson et al.\ 1998).

\subsection{Selection Criteria}

\begin{figure*}
\epsscale{1.18}
\plotone{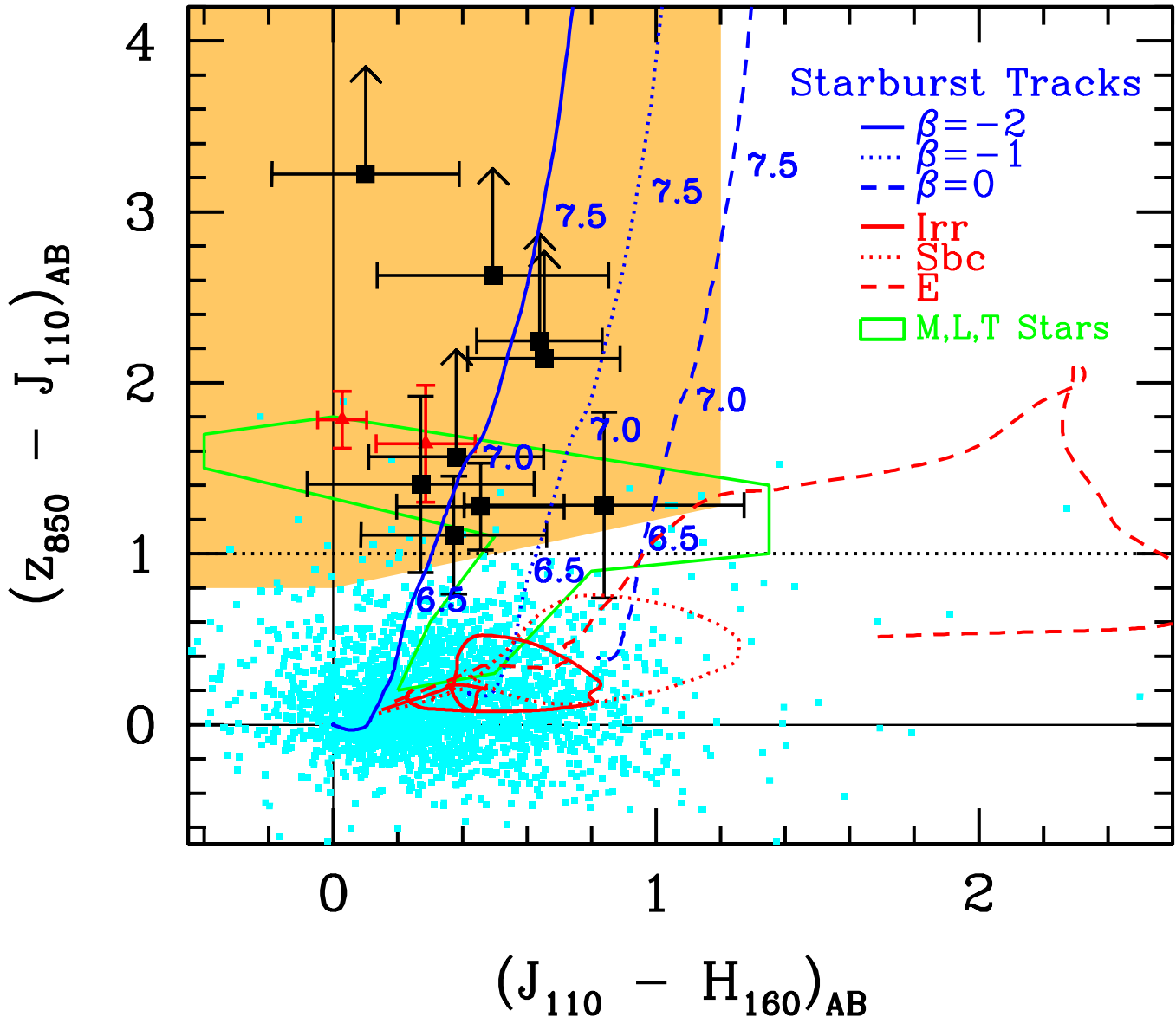}
\caption{$z-J$ and $J-H$ color-color diagram illustrating our
optical-IR $z$-dropout selection window (orange region).  The large
black squares show the colors of $z$-dropouts in our selection (as
well as one $z$-dropout found behind Abell 1689: Bradley et al.\ 2008)
and their $1\sigma$ errors (or lower limits).  Sources which do not
satisfy our $z$-dropout selection criteria are shown in cyan.  The
solid, dotted, and dashed blue lines show how the colors of starbursts
with $UV$-continuum slopes $\beta$ of $-2$, $-1$, and 0, respectively,
vary as a function of redshift.  Typical $UV$-continuum slopes $\beta$
observed for star-forming galaxies at $z\sim5-6$ are $-2$ (e.g.,
Stanway et al.\ 2005; Bouwens et al.\ 2006; Yan et al.\ 2005).  The
red lines show how these colors vary as a function of redshift for
low-redshift galaxies (Coleman, Wu, \& Weedman 1980).  The green lines
bracket the region in color-color space where we expect T dwarfs to
lie (Knapp et al.\ 2004).  The two red triangles indicate the colors
of two bright T dwarfs found in our search (\S3.3).  While five of the
z-dropout candidates in our selection nominally have $z-J$ and $J-H$
colors consistent with that of T-dwarfs, their $z$-band flux appears
to be extended in all five cases (see also \S3.2-\S3.4), so these
sources cannot be stars.\label{fig:zjjh}}
\end{figure*}

Our selection criterion for our $z_{850}$-dropout search was
$((z_{850}-J_{110})_{AB})>0.8)\wedge
((z_{850}-J_{110})_{AB}>0.8+0.4(J_{110}-H_{160})_{AB})$ as shown in
Figure~\ref{fig:zjjh}, where $\wedge$ represents the logical
\textbf{AND} operation.  This criterion is identical to the ``less
conservative'' $z$-dropout selection criterion of Bouwens \&
Illingworth (2006).  We opted to use this criterion rather the more
stringent ``conservative'' criterion from that same work for two
reasons.  First, we realized that our ``conservative'' criterion may
have been more stringent than necessary to reject contaminants like
T-dwarfs (low mass stars) from our samples.  These contaminants can
(in most cases) be rejected by a procedure we will describe below.
Second, by adopting a ``less conservative'' selection criteria, the
effective selection volumes will be larger and therefore less
sensitive to small inaccuracies in the assumed model galaxy
population.

For regions of the two GOODS fields with ISAAC/MOIRCS coverage, our
$z$-dropout criterion was $((z_{850}-J)_{AB}>1.0)\wedge(J-K_{s}<1.2)$.
This selection criterion is very similar to the criterion used over
our NICMOS fields.  For the typical star-forming galaxy at $z\sim7$
(with a $UV$-continuum slope $\beta$ of $-2$) this criterion selects
galaxies to a slightly lower-redshift limit $z\gtrsim6.5$ than our
NICMOS $z$-dropout criterion (where $z\gtrsim6.6$).

The criterion for our $z\sim9$ $J_{110}$-dropout search was
$(J_{110}-H_{160})_{AB}>1.3$.  We elected to use a less restrictive
$(J_{110}-H_{160})_{AB}$ color criterion than the
$(J_{110}-H_{160})_{AB}>1.8$ criterion used in our previous
$J$-dropout search (Bouwens et al.\ 2005) because of the availability
of deep optical data over almost all of our previous search fields to
eliminate low redshift interlopers.  Previously ultra deep optical
data were not available over the NICMOS parallels to the HUDF, and so
we accordingly decided to select $z$ dropouts using a more stringent
$J-H$ color cut.

In computing the $z-J$ and $J-H$ colors that we used for our $z$ and
$J$ dropout selections, we set the $z$- and $J$-band fluxes to their
$1\sigma$ upper limits in the case of non-detections.  This was to
minimize the extent to which faint sources in our search fields would
scatter into our dropout samples.

Sources in our $z$ and $J$-dropout selections were also required to be
detected at the $4.5\sigma$ level in the $H_{160}$ band within a
0.6$''$ diameter aperture (to eliminate spurious sources) and
undetected ($<2\sigma$) in the available $B_{435}$, $V_{606}$, or
$i_{775}$ band imaging.  Sources which were detected at the
$1.5\sigma$ level in more than one of the ACS $B_{435}$, $V_{606}$,
and $i_{775}$ bands were also rejected as low-redshift interlopers.
Possible contamination by low mass stars (which are a concern for our
$z$-dropout selection) are identified by looking for sources which
showed faint detections ($>2\sigma$) in the $z_{850}$-band in a
$0.2''$-diameter aperture and whose $z_{850}$-band profile appeared to
be consistent with a point source (i.e., the SExtractor stellarity
parameter $>0.6$).

All candidate dropouts were inspected in the available IRAC and MIPS
coverage over the two GOODS fields (Dickinson \& GOODS team 2004) to
ensure that none had particularly red $3.6\mu-5.8\mu\gtrsim1$ mag or
$3.6\mu-24\mu\gtrsim2$ mag colors.  With the exception of the two
sources which satisfy our $J$-dropout criterion (03:32:41.22,
$-27$:44:00.8 and 03:32:32.45, $-27$:42:21.5), none of the sources
which satisfied our other criteria were found to have such red
infrared colors (or showed significant detections at $5.8\mu$,
$8.0\mu$, or $24\mu$).

The two particularly red $J_{110}$ dropouts (with $H_{160}$ magnitudes
of $25.7\pm0.2$ and $24.5\pm0.1$, respectively) have
$H_{160}-[4.5\mu]$ colors of $\sim2.5$ mag and are even detected at
$24\mu$ in the available MIPS imaging over the GOODS fields (meaning
that their apparent magnitudes at $24\mu$m are $\lesssim21.5$ mag).
Since we do not expect galaxies at $z\gtrsim7$ to be so bright at
$24\mu$ and otherwise so highly reddened (but see also discussion in
Wiklind et al.\ 2007), we have excluded these sources as candidate
$z\sim9$ galaxies.

\begin{figure}
\epsscale{1.20}
\plotone{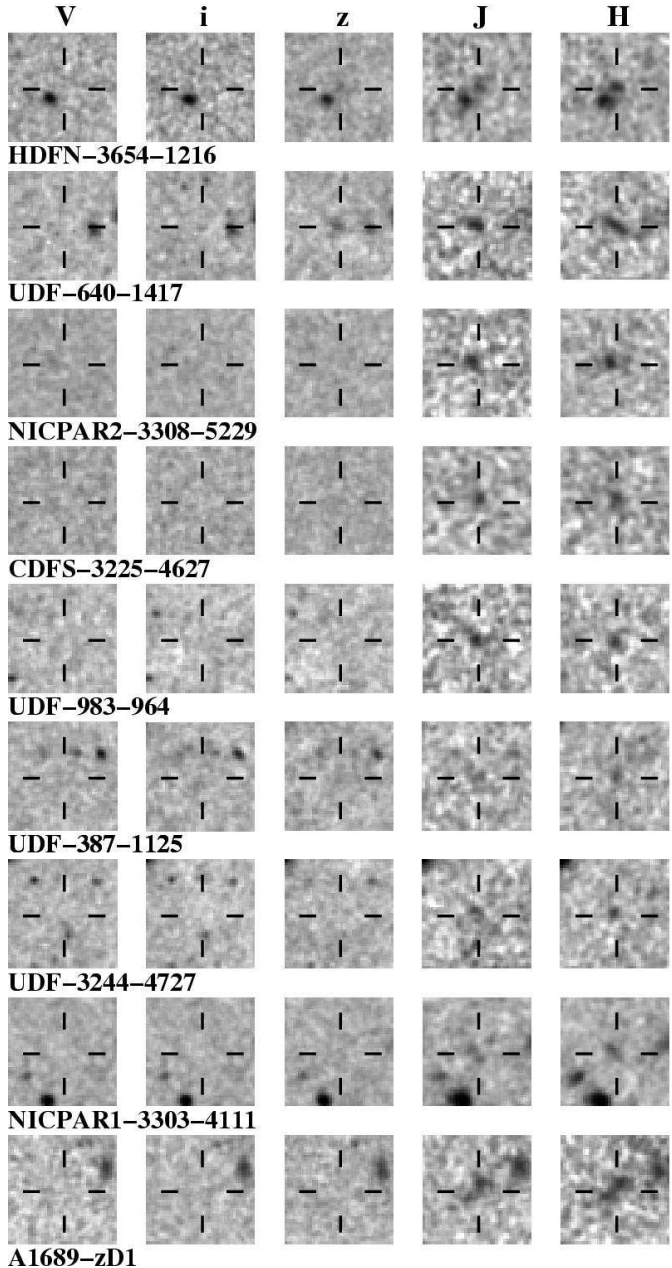}
\caption{$V_{606}$, $i_{775}$, $z_{850}$, $J_{110}$, and $H_{160}$
band cutouts for all eight sources in our $z\sim7$ $z_{850}$-dropout
sample.  The source presented in the bottom row is another very strong
$z\sim7$ $z$-dropout candidate found in an independent search around
massive lensing clusters (Bradley et al.\ 2008; Bouwens et al.\ 2008).
All the sources in our sample are detected at $>4.5\sigma$ in both the
$J_{110}$ and $H_{160}$ bands, with the exception of UDF-3244-4727 and
UDF-387-1125 which is only detected at $3\sigma$ in the $J_{110}$ band
(we will be obtaining deeper $J_{110}$-band imaging data for these
latter two sources as part of an approved HST program).  All the
sources in our sample show strong $z-J$ breaks and are completely
undetected ($<2\sigma$) in the $V_{606}$ and $i_{775}$ bands (see
\S3.2).
\label{fig:zstamp}}
\end{figure}

\subsection{Selection Results}

Application of our selection criteria across all the fields given in
Table~\ref{tab:obsdata} yielded 8 credible $z_{850}$-dropout
candidates, but no credible $J_{110}$-dropout candidates.  All 8 of
our $z$-dropout candidates were found in our NICMOS search fields.
The properties of each of the 8 $z_{850}$-dropout candidates are given
in Table~\ref{tab:zcandlist}.  Figure~\ref{fig:zstamp} provides a
postage stamp montage of the 8 candidates.  We also included one
candidate we identified in a separate search around massive lensing
clusters (Bradley et al.\ 2008; Bouwens et al.\ 2008).  Five of the
eight candidates were given in our two previously published
$z_{850}$-dropout searches (Bouwens et al.\ 2004c; Bouwens \&
Illingworth 2006).  Three of these candidates are new.  Two of the
three new candidates (UDF-3244-4727 and NICPAR1-3303-4111) were just
below the flux thresholds used in our previous searches, but meet
these thresholds now as a result of modest improvements in the S/N and
overall quality of our NICMOS reductions. The third candidate
(NICPAR2-3308-5229) was found over the second NICMOS parallel to the
HUDF and could be identified as a result of the much deeper optical
(ACS) data now available over these fields to set strong constraints
on the $z-J$ colors.  Our sample also includes one $z_{850}$-dropout
(UDF-380-1125) which, while appearing in the our initial $z$-dropout
selection over the HUDF (Bouwens et al. 2004c), did not appear in our
more recent Bouwens \& Illingworth (2006) selection because of the
more conservative S/N threshold used in that work.

\begin{deluxetable*}{cccccccc}
\tablecolumns{8} \tablecaption{$z_{850}$-dropout
candidates.\label{tab:zcandlist}} \tablehead{ \colhead{Object ID} &
\colhead{R.A.} & \colhead{Dec} & \colhead{$H_{160}$} &
\colhead{$z_{850}-J_{110}$} & \colhead{$J_{110}-H_{160}$} &
\colhead{$H_{160}-K_s$} & \colhead{Ref\tablenotemark{a}}} \startdata

HDFN-3654-1216 & 12:36:54.12 & 62:12:16.2 & $26.0\pm0.1$ & $1.1\pm0.3$
& $0.4\pm0.3$ & $0.0\pm0.3$ & [2] \\ 

UDF-640-1417 & 03:32:42.56 &
$-$27:46:56.6 & $26.2\pm0.1$ & $1.3\pm0.3$ & $0.5\pm0.2$ & $0.5\pm0.3$
& [1,2,3,4,5] \\ 

NICPAR2-3308-5229\tablenotemark{b} & 03:33:08.29 &
$-$27:52:29.2 & $26.7\pm0.1$ & $>2.2$\tablenotemark{c} & $0.6\pm0.2$ &
--- & --- \\ 

CDFS-3225-4627 & 03:32:25.22 & $-$27:46:26.7 &
$26.7\pm0.2$ & $1.4\pm0.5$ & $0.2\pm0.3$ & $<0.2$\tablenotemark{c} & [2] \\ 

UDF-983-964 &
03:32:38.80 & $-$27:47:07.2 & $26.9\pm0.2$ & $>3.2$\tablenotemark{c} &
$0.1\pm0.3$ & $0.0\pm0.8$ & [1,2,4,5] \\ 

UDF-387-1125 & 03:32:42.56 & $-$27:47:31.4 & $27.1\pm0.2$ &
$1.3\pm0.5$ & $0.8\pm0.4$ & $<0.0$\tablenotemark{c} & [1,5] \\

UDF-3244-4727 & 03:32:44.02 &
$-$27:47:27.3 & $27.3\pm0.2$ & $>2.6$\tablenotemark{c} & $0.5\pm0.4$ &
$<0.1$\tablenotemark{c} & --- \\ 

NICPAR1-3303-4111\tablenotemark{d} &
03:33:03.81 & $-$27:41:12.1 & $27.8\pm0.1$ & $>1.5$\tablenotemark{c} &
$0.4\pm0.2$ & --- & --- \\

\multicolumn{7}{c}{------------------------------------------------------------------------------}\\
A1689-zD1\tablenotemark{e} & 13:11:29.73 & $-$01:19:20.9 &
$24.7\pm0.1$ & $>2.2$\tablenotemark{c} & $0.6\pm0.2$ & --- & [6]
\enddata \tablenotetext{a}{References: [1] Bouwens et al.\ (2004c),
[2] Bouwens \& Illingworth (2006), [3] Yan \& Windhorst (2004), [4]
Coe et al.\ (2006), [5] Labb\'{e} et al.\ (2006), [6] Bradley et al.\
(2008)}

\tablenotetext{b}{There appears to be some flux in the $V_{606}$ band
very close to the position of this source ($\sim0.1''$).  If this flux
is associated with this source and not a chance superposition, it
would rule out this source being a $z\gtrsim7$ galaxy.  The existence
of a small amount of flux in the $V_{606}$ band would be somewhat
unexpected given that there is no detectable flux in the $i_{775}$ and
$z_{850}$ bands and the fact that these exposures reach similar flux
limits.  On balance, we consider this to be a likely $z\sim7$
candidate.}
\tablenotetext{c}{Upper and lower limits on the measured colors are
the $1\sigma$ limits.}
\tablenotetext{d}{Because this source (detected at $\sim5-6\sigma$)
was found over the NICMOS parallels to the HUDF and the noise
properties of these data may exhibit significant non-Gaussian
characteristics due to the small number of unique dither positions
(see \S2.1), we took special care in assessing the reality of this
source.  We considered all the available NICMOS data at 4 highly
distinct dither positions (separated by $\sim2-3''$ each) and in both
the $J_{110}$ and $H_{160}$-band data separately.  The candidate was
evident at $\gtrsim2-3\sigma$ significance in 7 of the 8 of the
stacks, consistent with what we might expect if the candidate were a
real source.  This source is also present in an independent reduction
of these data (Oesch et al.\ 2008).}

\tablenotetext{e}{This source was identified in a separate search for
$z\sim7$ $z$-dropouts behind massive galaxy clusters (Bradley et al.\
2008; Bouwens et al.\ 2008) and is not used here for deriving the $UV$
LF at $z\sim7$.  This source is substantially magnified by
gravitational lensing from Abell 1689 ($\mu\sim9.3$).  Its delensed
$H_{160}$-band magnitude is $\sim27.1$ AB mag (Bradley et al.\ 2008).}
\end{deluxetable*}

There were two sources in our catalogues which while satisfying our
$z$-dropout color criteria did not make it into our $z$-dropout
selection because of their compactness in the $z$-band, which strongly
suggested they were T dwarfs (03:32:25.10, $-27$:46:35.6 and
12:37:34.26, 62:18:31.4).  These sources had previously been
identified as possible T dwarfs in a number of other studies (Bouwens
\& Illingworth 2006; Mannucci et al.\ 2007; Eyles et al.\ 2007).
There were also 28 sources in our selection that while satisfying our
$z$-dropout color criteria showed significant detections ($>2\sigma$)
in the $B_{435}$, $V_{606}$, or $i_{775}$ bands and thus were excluded
from our $z$-dropout selection.  One possible $z$-dropout candidate
over the near-IR ISAAC data was found at 03:32:24.74, $-27$:55:15.8.
This source seemed to be a plausible $z\sim7$ candidate, with an
apparent $5\sigma$ detection in the $J$ band, an apparent $J$-band
magnitude of 24.8$\pm$0.2 AB mag, and no detection ($<1\sigma$) in any
of the optical imaging data.  However, since this source is undetected
($H_{AB} >26.0$ [$3\sigma$ limit] in a $1''$-diameter aperture) in
even deeper NICMOS $H_{160}$-band data over this field, it seems
highly likely that this source is spurious (or perhaps a SNe: see
\S3.4).


In Appendix D, we discuss several sources which appeared in previous
$z\gtrsim7$ dropout samples of ours but do not appear in the current
selections.

\subsection{Contamination}

We give particular attention to contamination, since it can be a
significant issue in searches for very faint, high-redshift galaxies.
This discussion is only relevant for our $z_{850}$-dropout selection
since our $J_{110}$-dropout selection contains no credible candidates.
There are four types of contaminants that are of particular concern:
\textit{(i)} T-dwarfs, \textit{(ii)} supernovae, \textit{(iii)}
lower-redshift galaxies which may move into our $z_{850}$-dropout
selection through photometric scatter, and \textit{(iv)} spurious
sources.

\textit{Possible Contamination from T dwarfs:} T dwarfs are low-mass
brown dwarfs, which while most well known and studied in the solar
neighborhood (e.g., Knapp et al.\ 2004; Burgasser et al.\ 2006), are
sometimes observed at great distances and therefore can be quite faint
(i.e., $J_{AB}\sim26$: Ryan et al.\ 2005; Bouwens \& Illingworth
2006).  Since T dwarfs occupy a position in color-color space very
similar to that of high-redshift galaxies (Figure~\ref{fig:zjjh}),
they can therefore be difficult to distinguish from star-forming
galaxies at $z\sim7$ on the basis their photometry alone.  This leaves
us with the size or morphology of these sources as essentially our
only means of distinguishing these sources from $z\sim7$ galaxies.
This enterprise is made more difficult because the T dwarfs are
brightest in exactly the same data (near-IR) where the resolution is
the poorest.  Fortunately, the available ACS data have sufficient
depth and resolution that any T dwarfs in our data should show up as
$\gtrsim5\sigma$ point sources.  This follows from the fact that all
candidate dropouts in our sample are detected at $\gtrsim4.5\sigma$, T
dwarfs have $z-J$ colors of $\sim1-2$ mag, and the ACS $z_{850}$-band
data are at least $\sim2$ mag deeper in a $0.2''$-diameter aperture
(ACS resolution) than our near-IR data is in a $0.6''$-diameter
aperture (near-IR resolution: see
Table~\ref{tab:obsdata}).\footnote{The $5\sigma$ depths of the ACS
data in a $0.2''$-diameter aperture is $\sim1.2$ mag deeper than that
given in Table~\ref{tab:obsdata} (for a $0.6''$-diameter aperture).}
Since none of the $z$-dropouts in our selection show
$\gtrsim4\sigma$-pointlike detections in the available $z_{850}$-band
coverage, this suggests that these sources are not T dwarfs.

\textit{Possible Contamination from Supernovae:} Since the NICMOS $JH$
data were largely taken at quite different times than the ACS $BViz$
data, it is quite possible that highly time-variable sources like
supernovae could show up as dropouts in our selections.  To estimate
the approximate frequency at which we might expect such sources to
show up in our searches, we adopted the GOODS supernova searches
(Riess et al.\ 2004; Strolger et al.\ 2004) as a baseline.  In their
searches, they identified 42 supernovae in 8 independent searches over
one of the two $\sim160$ arcmin$^2$ GOODS fields (15 ACS WFC
pointings), or $\sim1/3$ supernovae per $\sim11$ arcmin$^2$ ACS WFC
field per $\sim40$ day period.  Of these supernovae, only $\sim40$\%
of the high-redshift supernovae identified by Riess et al.\ (2004) and
Strolger et al.\ (2004) appear to be well enough separated from their
host galaxies to be identified as a separate source.  Having
calculated the approximate rate at which supernovae appear at random
in deep fields, all that remains is to calculate the time period over
which we might expect these supernovae to show up as possible sources
in our search.  While theoretically we might expect high-redshift SNe
to be bright enough to show up in our survey data for $\sim100$ days
(if we take the typical light curves found in e.g., Strolger et al.\
2004), all $z$-dropouts in our sample are within $\sim1$ mag of the
flux limit of each survey field.  Typical SNe (at $z\sim1$) would not
remain at these magnitudes for more than $\sim40$ days, so the
incidence of SNe in our probe should be $\sim(1/3)(1/11
\textrm{arcmin}^2)(0.40)\sim0.012\, \textrm{arcmin}^{-2}$ for data
taken at different epochs.  Multiplying this by the $\sim16$
arcmin$^2$ of NICMOS data in our search taken at significantly
different times than the ACS data (the HUDF NICMOS data were taken at
a similar time to the ACS HUDF data), we would expect that only
$\sim$0.2 supernovae should contaminate our high redshift selections.
While this number is just 3\% of the 8 sources in our $z$-dropout
selection and therefore likely of minimal importance for the present
LF determinations, it does indicate that contamination of
high-redshift selections by supernovae is a real possibility.  This
may be a particularly important concern in searches for bright
$z\gtrsim7$ galaxies due to the inherent rarity of those sources.

\textit{Possible Contamination Through Photometric Scatter:} One
possible source of contamination are low redshift sources scattering
into our color selection simply because of noise.  To assess the
importance of this effect, we ran a number of Monte-Carlo simulations
where we started with a low-redshift galaxy population and then added
noise to the fluxes in the individual bands to see how many of the
sources would satisfy our $z_{850}$-dropout selection.  Clearly, the
most critical aspect of these simulations is ensuring that the
properties of the input galaxy population (to which noise is added)
are accurate.  We set up the input population as follows: (1) The
input $H_{160,AB}$ magnitudes and photometric errors are modelled
using the actual distribution of magnitudes and errors found in the
data (to make our estimates as model independent as possible and
because essentially all faint sources in our fields will be at
$z<6.5$) and (2) The colors (i.e., optical-$H_{160}$ and
$J_{110}-H_{160}$) of the input population are assumed to be
distributed in the same way as is observed for galaxies in the range
$24.5<H_{160,AB}<26$ in the HUDF (ACS+NICMOS coverage).  We used
galaxies in this magnitude range because of their higher S/N colors
and because the multivariate color distribution is similar (albeit
slightly redder) to galaxies at $H_{160}$-band magnitudes $\gtrsim26$
AB mag.  This magnitude range also does not include any sources which
are candidate $z\sim7$ galaxies (at least in the HUDF).

Based upon the described input population, we ran 10 separate
photometric scattering experiments for each search field.  We
estimated that there would be 0.0, 0.3, 0.2, and 0.3 low-redshift
contaminants, respectively, for our $z$-dropout selections over the
HUDF, NICPAR, HDFN, and GOODS parallel fields.  In total, $\sim0.8$
contaminants are expected from all our fields.  We note that the
approach taken here to estimate contamination from photometric scatter
is analogous to the one described in Bouwens et al.\ (2004c) or
Appendix D.4.2 of Bouwens et al.\ (2006).

\textit{Possible Contamination from Spurious Sources:} Finally, to
estimate the number of contaminants from spurious sources, we repeated
our selection procedure on the negative images (e.g., Dickinson et
al.\ 2004; Yan \& Windhorst 2004; Bouwens et al.\ 2007) and found no
sources that satisfied our selection.  This suggests that the spurious
fraction is negligible.  We note that we would not expect
contamination from spurious sources to be a problem since essentially
all of the sources in our sample are detected at $\geq4.5\sigma$ in
both the $J_{110}$ and $H_{160}$ bands.  The only notable exceptions
to this are for the two sources in our sample, UDF-387-1125 and
UDF-3244-4727, where $J_{110}$-band detections are still only at the
$3\sigma$ level.  However, in both cases, we are attempting to improve
the detection significance (as well as the S/N of the $z-J$ and $J-H$
colors) by obtaining an additional 9-12 orbits of $J_{110}$-band data
on both sources (likely improving their detection significance to the
$4.5\sigma$ level).

In summary, the overall level of contamination should be fairly low in
general.  We would expect $\sim0.8$ galaxy to enter our sample from
photometric scatter, $\sim0.2$ SNe contaminants to enter our sample,
and there to be no contamination from T dwarfs or spurious sources.
Overall, this suggests an overall contamination level of $\sim12$\%.

\section{Determinations of the UV LF at $z\sim7-10$}

We now utilize our search results to determine the rest-frame $UV$ LF
at $z\sim7$.  We will consider both LFs using both stepwise and
Schechter parametrizations.  The advantage of the stepwise
determinations is that they give a more model-independent measure of
the constraints we have on the LF, while the Schechter determinations
are more amenable to direct interpretation and comparison with other
determinations in the literature.

As in other methodologies for determining the LF, we maximize the
probability with which our model LFs are able to reproduce our search
results.  Since we do not know the precise redshifts for our sources,
our goal will simply be to reproduce the surface density of
$z$-dropout candidates in each of our search fields as a function of
apparent magnitude.  This is similar to the modelling we performed in
our previous paper on the $UV$ LFs at $z\sim4$, $z\sim5$, and $z\sim6$
(Bouwens et al.\ 2007), where we attempted to reproduce the surface
density of $B$, $V$, and $i$ dropouts in deep ACS fields.

To maximize the likelihood with which our model LFs reproduce our
search results, we need to be able to compute the expected surface
density of dropouts as a function of magnitude.  We use a similar
formalism to the one we used in our previous paper on the LFs at
$z\sim4-6$:
\begin{equation}
\Sigma _{k} \phi_k V_{m,k} = N_m
\label{eq:numcountg}
\end{equation}
where $N_m$ is the surface density of galaxies in some search field
with magnitude $m$, $\phi_k$ is the volume density of galaxies with
absolute magnitude $k$, and $V_{m,k}$ is the effective selection
volume for which galaxies with absolute magnitude $k$ will both
satisfy our dropout selection criteria and be observed to have an
apparent magnitude $m$.  The width of the magnitude bins in $N_m$,
$\phi_k$, and $V_{m,k}$ will depend upon the LF parameterization we
consider (stepwise vs. Schechter).

We computed the effective volumes $V_{m,k}$ for our $z$ and $J$
dropout selections by artificially redshifting $B$-dropouts from the
HUDF across the selection windows of our higher redshift samples,
adding the sources to the real data, and then repeating our selection
procedure.  We used our well-tested cloning software (Bouwens et al.\
1998a,b; Bouwens et al.\ 2003) to handle the artificial redshifting of
sources from $z\sim4$ to $z\sim6-10$.  The size of the sources were
scaled as $(1+z)^{-1.1}$ (independent of luminosity) to match the size
trends observed for star-forming galaxies at $z\gtrsim2$ (Bouwens et
al.\ 2006; see also Ferguson et al.\ 2004 and Bouwens et al.\ 2004c).
The redshifted (``cloned'') sources were smoothed so that their
effective PSFs matched that seen in the data.  The mean $UV$-continuum
slope $\beta$ was set to $-2$ to agree with that observed in
$L_{z=3}^{*}$ star-forming galaxies at $z\sim6$ (Stanway et al.\ 2005;
Bouwens et al.\ 2006; Yan et al.\ 2005) while the $1\sigma$ dispersion
in $\beta$ was set to 0.5, which is similar to what is found at
$z\sim4-6$ (e.g., Bouwens et al.\ 2007; R.J. Bouwens et al.\ 2008, in
prep).  Instead of using simple power laws to represent model SEDs of
given $UV$ continuum slope $\beta$, we elected to use $10^8$-yr
continuous star-formation models (Bruzual \& Charlot 2003) where the
dust extinction (Calzetti et al.\ 1994) is varied to reproduce the
model slopes.  This should provide for a slightly more realistic
representation of the SEDs of star-forming galaxies at $z\gtrsim7$
than can be obtained from simple power law spectra.

To evaluate the likelihood with which the surface densities computed
above would reproduce that found in the observations, we assumed
simple Poissonian statistics and compared the observed and expected
numbers bin by bin.  In detail, our expression for the likelihood
$\cal{L}$ is
\begin{equation}
{\cal L}=\Pi_{i,j} e^{-N_{exp,i,j}} \frac{(N_{exp,i,j})^{N_{obs,i,j}}}{(N_{obs,i,j})!}
\label{eq:ml}
\end{equation}
where $N_{obs,i,j}$ is the observed number of sources in search field
$i$ and magnitude interval $j$, where $N_{exp,i,j}$ is the expected
number of sources in search field $i$ and magnitude interval $j$, and
where $\Pi_{i,j}$ is the product symbol.  We elected to use the above
procedure instead of a more conventional procedure like STY79
(Sandage, Tammann, \& Yahil 1979) or SWML (Efstathiou et al.\ 1988:
which determine the shape of the LF independent of the normalization)
to take advantage of our search constraints on the volume density of
bright $z\gtrsim7$ galaxies from our wide-area ground-based data.
Maximum likelihood procedures (like SWML or STY79) do not take
advantage of these null search results -- simply because there are no
sources to use for estimating the shape of the LF.  While the maximum
likelihood fit results from Eq.~\ref{eq:ml} will be affected by
large-scale structure variations across our search fields, we will
compute the uncertainties which result from these variations in \S4.1,
\S4.2, and Appendix A and include them in our overall error budget.

In determining the rest-frame UV LF at $z\sim7$ and $z\sim9$ from
our $z$ and $J$ dropout selections, we will adopt $z\sim7.3$ and
$z\sim9.0$, respectively, as the fiducial redshifts for these
selections and the derived LFs.  In \S4.2, we will justify the use of
these mean redshifts using the effective volume simulations just
described (see also Bouwens et al.\ 2004; Bouwens et al.\ 2005;
Bouwens \& Illingworth 2006).

\subsection{Stepwise Determinations}

We begin with a determination of the LF at $z\sim7$ and $z\sim9$ in
stepwise form.  The advantage of stepwise parametrizations is that
they allow us to determine the overall prevalence of galaxies in a way
that is very model independent and just a simple function of
luminosity.  The width of the bins in the LF are 0.8 mag and was
chosen as a compromise between the need to have a few $z\sim7$
galaxies per bin and the desire to retain a reasonable resolution in
luminosity for our derived LFs.  We use a similar bin size for the
stepwise upper limits on our $z\sim9$ $J$-dropout LF in order to make
these limits more easily interpretable.  The stepwise LF $\phi_k$ is
then derived using Eq.~\ref{eq:numcountg}.  The results are presented
in Table~\ref{tab:swlf}.

Perhaps the largest uncertainties in these LF determinations arise
from small number statistics (Poissonian errors).  However, in
addition to the simple number statistics, there are also sizeable
uncertainties that come from large-scale structure variations as well
as our effective volume estimates.  Let us first consider the
uncertainties coming from large-scale structure variations.  Adopting
a pencil beam geometry, redshift selection window with width $\Delta z
= 1.5$, search area of 100 arcmin$^2$, and bias of 7 (which is
appropriate for sources with volume densities of $\sim10^{-3.5}$
Mpc$^{-3}$: e.g., Mo \& White 1996; Somerville et al.\ 2004; Trenti \&
Stiavelli 2008), we estimate a $1\sigma$ RMS uncertainty of $\sim$30\%
in the $\phi_k$'s due to field-to-field variations (see also estimates
in Bouwens \& Illingworth 2006).  At $z\sim9$, these uncertainties
are $\sim20$\% assuming a redshift selection window with width $\Delta
z = 2$ (see Bouwens et al.\ 2005).

Uncertainties in our effective volume estimates derive primarily from
our imperfect knowledge of the size (or surface brightness)
distribution of star-forming galaxies at
$z\gtrsim7$.\footnote{Uncertainties in the UV color distribution also
contribute to the overall error budget for our effective volume
estimates, but they are smaller in general (e.g., see \S5.2).}
Fortunately, the mean size of star-forming galaxies at $z\gtrsim4$
show a good correlation with redshift (i.e., mean half-light radius
$\propto (1+z)^{-1.1\pm0.3}$ for fixed luminosity: Bouwens et al.\
2006) and we can make a reasonable estimate for what the size (surface
brightness) distribution of galaxies is at $z\gtrsim7$.  Nonetheless,
this distribution is at least as uncertain as the error on the
size-redshift scaling.  Propagating the error on this scaling into the
size distributions assumed in our effective volume estimates, we
estimate an RMS uncertainty of 17\% in the selection volume at
$z\sim7$ and 15\% at $z\sim9$ due to the uncertainties in the size
(surface brightness) distribution.  Together the size and large-scale
structure uncertainties add an uncertainty of 34\% and 25\% RMS to
each bin of the rest-frame UV LF at $z\sim7$ and $z\sim9$,
respectively.  These uncertainties have been added in quadrature with
those deriving from the small number statistics.  They are given in
Table~\ref{tab:swlf}.

\begin{deluxetable}{lcc}
\tablewidth{0pt}
\tabletypesize{\footnotesize}
\tablecaption{Stepwise Constraints on the rest-frame $UV$ LF at $z\sim7$ and $z\sim9$ (\S4.1).\label{tab:swlf}}
\tablehead{
\colhead{$M_{UV,AB}$\tablenotemark{b}} & \colhead{$\phi_k$ (Mpc$^{-3}$ mag$^{-1}$)}}
\startdata
\multicolumn{2}{c}{$z$-dropouts ($z\sim7$)}\\
$-21.55$ & $<0.000003$\tablenotemark{a} \\
$-20.84$ & $0.00005_{-0.00003}^{+0.00007}$\\
$-20.04$ & $0.00029_{-0.00014}^{+0.00024}$\\
$-19.24$ & $0.00063_{-0.00041}^{+0.00085}$\\
\multicolumn{2}{c}{$J$-dropouts ($z\sim9$)}\\
$-21.95$ & $<0.000019$\tablenotemark{a}\\
$-21.15$ & $<0.000023$\tablenotemark{a}\\
$-20.35$ & $<0.000082$\tablenotemark{a}\\
$-19.55$ & $<0.000687$\tablenotemark{a}\\
\enddata 
\tablenotetext{a}{Upper limits here are $1\sigma$ (68\% confidence).}
\tablenotetext{b}{The effective rest-frame wavelength is $\sim1900\AA$
for our $z$-dropout selection and $\sim1500\AA$ for our $J$-dropout
selection.}
\end{deluxetable}

\begin{figure}
\epsscale{1.2}
\plotone{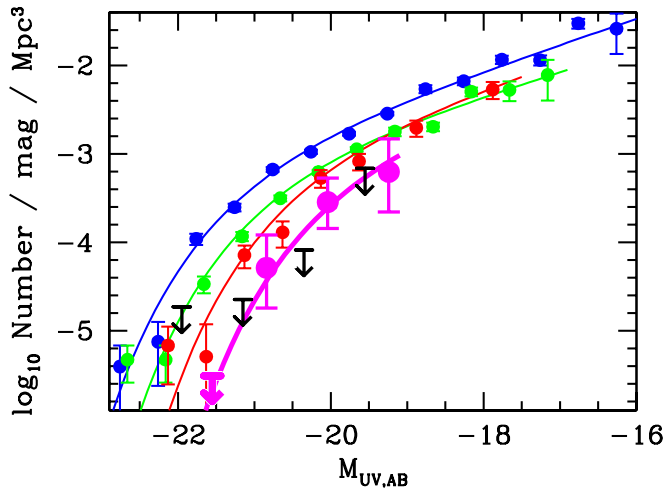}
\caption{Determinations of the rest-frame $UV$ luminosity function
(LF) at $z\sim7$ using both a Schechter parameterization
(\textit{magenta line}) and in stepwise form (\textit{magenta circles}
with $1\sigma$ error bars).  Note that the stepwise and Schechter
determinations of the LF are determined separately (i.e., our
Schechter LF fits are not obtained through fits to our stepwise LFs).
{\it The lines are not fits to the points.} The $1\sigma$ upper limits
on the bright end of the $UV$ LF at $z\sim7$ and at $z\sim9$ are
shown with the downward arrows in magenta and black, respectively.
For context, we have included the rest-frame $UV$ LFs determined by
B07 at $z\sim4$ (\textit{red symbols}), $z\sim5$ (\textit{green
symbols}), and $z\sim6$ (\textit{red symbols}).\label{fig:lf}}
\end{figure}

These LFs are also presented in Figure~\ref{fig:lf} with the magenta
points for our $z\sim7$ LF and black downward arrows for the
constraints on the $z\sim9$ LF.  A comparison with previous
determinations at $z\sim4$, $z\sim5$, and $z\sim6$ from Bouwens et
al.\ (2007) is also included on this figure for context.  Though the
error bars for individual points in the LF at $z\sim7$ are still
quite sizeable, there is strong evidence that the UV LF at $z\sim7$
is different from the UV LF at $z\sim6$ (99\% confidence) and thus
there is evolution from $z\sim7$ to $z\sim6$.  We determined this
confidence level by finding the value of $M^*$ and $\phi^*$ which
minimizes the total $\chi^2$ evaluated for our $i$ and $z$ dropout LFs
and then looking at the probability of obtaining the resultant
reduced-$\chi^2$ purely by chance.  This conclusion was already drawn
by Bouwens \& Illingworth (2006) on the basis of a smaller but very
similar selection of galaxies.

\begin{figure}
\epsscale{1.20}
\plotone{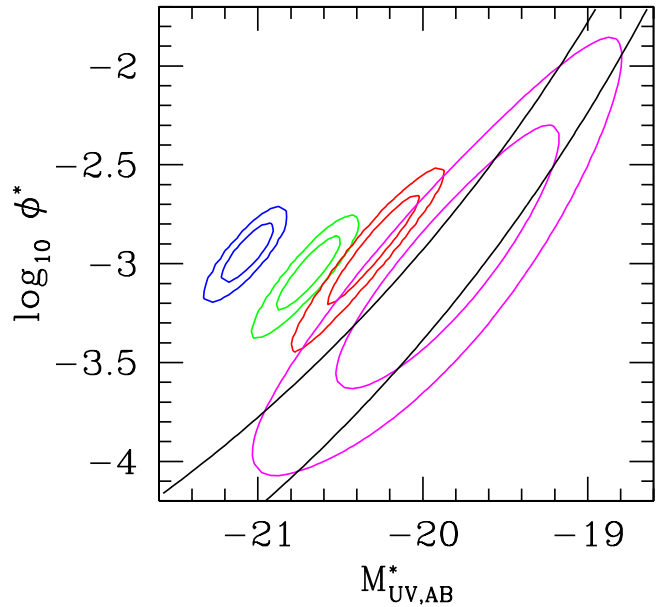}
\caption{68\% and 95\% likelihood intervals on the Schechter
parameters $M^*$ and $\phi^*$ for the rest-frame $UV$ luminosity
function (LF) at $z\sim7$ (\textit{magenta contours}) and $z\sim9$
(\textit{black contours}) derived from our $z$ and $J$-dropout search
results, respectively (where the left and right black lines represent
the 95\% and 68\% confidence intervals, respectively, on the $z\sim9$
LF).  Also shown are the likelihood contours Bouwens et al.\ (2007)
derived for the $UV$ LF at $z\sim4$ (\textit{blue contours}), $z\sim5$
(\textit{green contours}), and $z\sim6$ (\textit{red contours}).
These contours are expressed in terms of the Schechter parameters
$\phi^*$ and $M^*$.  A faint-end slope $\alpha$ of $-1.74$ is assumed
for the determinations at $z\sim7$ and $z\sim9$ (to match that seen
at $z\sim6$: Bouwens et al.\ 2007), but the exact value assumed for
$\alpha$ makes little difference to the derived contours for $\phi^*$
and $M^*$ (see \S4.2).  Not surprisingly, our constraints on the
Schechter parameters at $z\sim7$ are considerably worse than at
$z\sim4-6$, reflecting the substantially smaller size of samples at
$z\gtrsim7$.  Nonetheless, it seems clear that the $UV$ LF shows
significant and consistent evolution towards brighter values of $M^*$
with cosmic time (by $\sim1.2$ mag from $z\sim7$ to $z\sim4$).
There is little evidence for evolution in the volume density $\phi^*$.
Note that our constraints on the $z\sim9$ LF rule out no-evolution
from $z\sim9$ to $z\sim7$ at $\sim80$\% confidence.  No evolution
from $z\sim7$ to $z\sim6$ is ruled out at 99\% confidence.
\label{fig:contourml}}
\end{figure}

\subsection{Schechter Determinations}

We will now attempt to express the results of our search for
$z\gtrsim7$ galaxies using a Schechter parameterization.  The
Schechter parametrization is convenient since it is much more amenable
to interpretation than stepwise LFs are.  Of course, it is not at all
clear from the stepwise LFs derived in \S4.1 (particularly given the
sizeable observational uncertainties) that the $UV$ LF at $z\gtrsim7$
is well described by a Schechter function (see discussion in \S5.5).

As with our stepwise determinations, we will calculate the expected
surface density of dropouts given a model LF by using
Eq.~\ref{eq:numcountg} and expressing the Schechter function in
stepwise form.  For convenience, we have decided to bin the surface
density of galaxies in magnitude intervals of width 0.1 mag.  Use of
substantially finer bins does not have a noticeable effect on the
results.  Because of the size of current $z\sim7-10$ samples and
limited luminosities ($\lesssim-19$ AB mag) to which we can probe, we
cannot hope to obtain very strong constraints on the faint-end slope
of the LF at $z\gtrsim7$ and therefore it makes sense for us to fix it
to some fiducial value.  We will adopt $-1.74$, which is the faint-end
slope of the $UV$ LF at $z\sim6$ determined by Bouwens et al.\ (2007)
using the HUDF and a large number of deep ACS fields.  Later we will
investigate the sensitivity of our results to the value we assume for
the faint-end slope.

\begin{deluxetable}{ccccc}
\tablewidth{0pt}
\tabletypesize{\footnotesize}
\tablecaption{Comparison of the Schechter Parameters for
the rest-frame $UV$ LFs at $z\sim7$ and $z\sim9$ with
determinations at $z\sim4$, $z\sim5$, and $z\sim6$ from Bouwens
et al.\ (2007).\label{tab:lfparm}}
\tablehead{\colhead{Dropout} & \colhead{} & \colhead{} & \colhead{$\phi^*$
$(10^{-3}$} & \colhead{} \\
\colhead{Sample} & \colhead{$<z>$} &
\colhead{$M_{UV} ^{*}$\tablenotemark{a}} & \colhead{Mpc$^{-3}$)} &
\colhead{$\alpha$}}
\startdata
$z$ & 7.3 & $-19.8\pm0.4$ & $1.1_{-0.7}^{+1.7}$ & $(-1.74)$\tablenotemark{b}\\
$J$ & 9.0 & $>-19.6$\tablenotemark{c} & $(1.1)$\tablenotemark{b} &
$(-1.74)$\tablenotemark{b}\\
\multicolumn{5}{c}{-----------------------------------------------------------------} \\
$B$ & 3.8 & $-20.98\pm0.10$ & $1.3\pm0.2$ & $-1.73\pm0.05$\\ 
$V$ & 5.0 & $-20.64\pm0.13$ & $1.0\pm0.3$ & $-1.66\pm0.09$\\
$i$ & 5.9 & $-20.24\pm0.19$ & $1.4_{-0.4}^{+0.6}$ & $-1.74\pm0.16$
\enddata

\tablenotetext{a}{Values of $M_{UV}^{*}$ are at $1600\,\AA$ for the
Bouwens et al.\ (2007) $B$ and $V$-dropout LFs, at $\sim1350\,\AA$ for
the Bouwens et al.\ (2007) $i$-dropout LF, at $\sim1900\,\AA$ for our
$z$-dropout LF, and at $\sim1500\,\AA$ for our constraints on the
$J$-dropout LF.  Since $z\sim6$ galaxies are blue ($\beta\sim-2$:
Stanway et al.\ 2005; B06), we expect the value of $M_{UV}^*$ at $z\sim6$
to be very similar ($\lesssim0.1$ mag) at $1600\,\AA$ to its value at
$1350\,\AA$.  Similarly, we expect the value of $M_{UV}^*$ at $z\sim7$
and $z\sim9$ to be fairly similar at $\sim1600\AA$ to the values determined at
$\sim1900\AA$ and $\sim1500\AA$, respectively.}
\tablenotetext{b}{Fixed in our fits (\S4.2)} \tablenotetext{c}{Lower
limits here are $1\sigma$ (68\% confidence).}
\end{deluxetable}

\begin{figure}
\epsscale{1.20}
\plotone{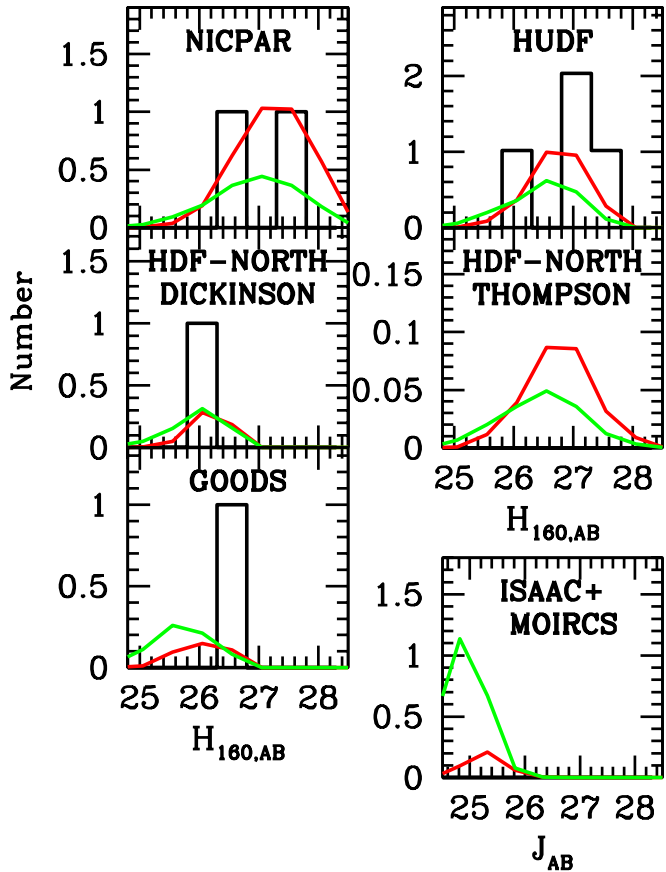}
\caption{Number of $z$-dropouts in our search fields (histogram)
versus that expected from our best-fit LF (red lines).  Panels
correspond to searches over the two NICMOS parallels to the HUDF
(Bouwens et al.\ 2005; Oesch et al.\ 2007), the NICMOS HUDF Thompson
\& Stiavelli fields (Thompson et al.\ 2005; Oesch et al.\ 2007), the
HDF-North Dickinson (1999) field, the HDF-North Thompson et al. (1999)
field, the CDF-South GOODS ISAAC field (B. Vandame et al.\ 2008, in
prep; J. Retzlaff et al.\ 2008, in prep: \S2.2), the HDF-North GOODS
MOIRCS fields (Kajisawa et al.\ 2006: \S2.3) and various NICMOS
parallels over the GOODS fields (see Table~\ref{tab:obsdata} and
Figure~\ref{fig:obsdata}).  Binning is done on 0.5 mag intervals.  The
green line shows the expected counts if we force our best-fit $UV$ LF
at $z\sim7$ to have the same value of $M_{UV}^*$ and $\alpha$ as we
determined at $z\sim4$ and only allow $\phi^*$ to vary to best fit the
data.  Clearly, the best-fit LF obtained in this manner (\textit{green
lines}) predicts too many galaxies at bright magnitudes than are
observed (e.g., $\sim2-3$ galaxies predicted at
$J_{AB},H_{160,AB}\sim25-26$ vs. the 0 observed) and far too few
galaxies at fainter magnitudes ($H_{160,AB}\sim27-28$: particularly in
the HUDF panel where the observed numbers are $>4\times$ larger).
This demonstrates that even in the very modest $z$-dropout selections
which are available there is reasonable ($\sim3\sigma$) evidence that
the typical luminosity of galaxies at $z\sim7$ is much fainter than it
is at $z\sim3-4$.  Of course, despite our ability to draw a few simple
conclusions like this, there are still significant uncertainties in
deriving the UV LF from the observed counts.\label{fig:expcount}}
\end{figure}

Our best-fit Schechter parameters at $z\sim7$ for our $z$-dropout
selection are $M_{AB} ^* = -19.8\pm0.4$ mag and
$\phi^*=0.0011_{-0.0007}^{+0.0017}$ Mpc$^{-3}$ for a fixed faint-end
slope $\alpha=-1.74$.  The 68\% and 95\% likelihood contours for these
parameters are given in Figure~\ref{fig:contourml} and compared with
our previous determinations from our $B$, $V$, and $i$-dropout
selections at $z\sim4$, $z\sim5$, and $z\sim6$, respectively
(Bouwens et al.\ 2007).  The best-fit values are also given in
Table~\ref{tab:lfparm}.  Large-scale structure uncertainties resulting
from field-to-field variations were estimated using Monte-Carlo
simulations (Appendix A) and incorporated into the uncertainties
quoted above.  While the best-fit value for $\phi^*$ is very similar
to that found at $z\sim4$ for the Bouwens et al.\ (2007) $B$-dropout
selections, the best-fit value for $M_{UV}^*$ is $1.2\pm0.4$ mag
fainter than the value of $M_{UV}^*$ (=$-20.98\pm0.07$ mag) found at
$z\sim4$ by Bouwens et al.\ (2007) and $0.4\pm0.4$ mag fainter than
the value of $M_{UV}^*$ ($=-20.24\pm0.19$ mag) found at $z\sim6$ by
Bouwens et al.\ (2007).  This suggests that the brightening we observe
in $M_{UV}^*$ from $z\sim6$ to $z\sim4$ (Bouwens et al.\ 2006; Bouwens
et al.\ 2007) is also seen from $z\sim7$.  Of course, we must admit
that we are somewhat surprised that our best-fit Schechter parameters
are in such excellent agreement with an extrapolation of lower
redshift trends!  It would suggest that our $z$-dropout sample may
largely be made up of star-forming galaxies at $z\sim7$ as we have
argued in \S 3.3 and \S3.4 (i.e., the number of low-redshift
interlopers is small) and that the effective volumes we have estimated
for this sample are reasonably accurate (see also discussion in
Appendix B).

\begin{deluxetable}{ccccc}
\tablewidth{0pt}
\tabletypesize{\footnotesize}
\tablecaption{Best-fit schechter parameters for the $UV$ LFs at $z\sim7$ and $z\sim9$ with different assumed faint-end slopes.\label{tab:lfparmalpha}}
\tablehead{\colhead{Dropout} & \colhead{} & \colhead{} & \colhead{$\phi^*$
$(10^{-3}$} & \colhead{} \\
\colhead{Sample} & \colhead{$<z>$} &
\colhead{$M_{UV} ^{*}$\tablenotemark{a}} & \colhead{Mpc$^{-3}$)} &
\colhead{$\alpha$}}
\startdata
$z$ & 7.3 & $-19.8\pm0.4$ & $1.1_{-0.7}^{+1.7}$ & $(-1.74)$\tablenotemark{b}\\
$z$ & 7.3 & $-19.6\pm0.4$ & $1.7_{-1.1}^{+2.6}$ & $(-1.4)$\tablenotemark{b}\\
$z$ & 7.3 & $-19.9\pm0.4$ & $0.8_{-0.5}^{+1.2}$ & $(-2.0)$\tablenotemark{b}\\
$J$ & 9.0 & $>-19.6$\tablenotemark{c} & $(1.1)$\tablenotemark{b} &
$(-1.74)$\tablenotemark{b}\\
$J$ & 9.0 & $>-19.5$\tablenotemark{c} & $(1.1)$\tablenotemark{b} &
$(-1.4)$\tablenotemark{b}\\
$J$ & 9.0 & $>-19.7$\tablenotemark{c} & $(1.1)$\tablenotemark{b} &
$(-2.0)$\tablenotemark{b}\\
$J$ & 9.0 & $(-21.0)$\tablenotemark{b} & $<0.06$\tablenotemark{c} &
$(-1.74)$\tablenotemark{b}\\
$J$ & 9.0 & $(-21.0)$\tablenotemark{b} & $<0.05$\tablenotemark{c} &
$(-1.4)$\tablenotemark{b}\\
$J$ & 9.0 & $(-21.0)$\tablenotemark{b} & $<0.07$\tablenotemark{c} &
$(-2.0)$\tablenotemark{b}\\
\enddata
\tablenotetext{a}{See remarks in Table~\ref{tab:lfparm}.}
\tablenotetext{b}{Fixed in our fits (\S4.2)}
\tablenotetext{c}{Lower limits here are $1\sigma$ (68\% confidence).}
\end{deluxetable}

Given the small size of current $z$-dropout samples, it may seem
surprising that we are able to obtain any constraint at all on the
shape on the $UV$ LF at $z\sim7$.  Fortunately, the large luminosity
range over which we have constraints on the surface density of
dropouts (i.e., from 25 AB mag to 28 AB mag) largely makes up for what
we lack in statistics.  These constraints can be helpful, even
brightward of 26.0 AB mag where our $z$-dropout sample contains no
sources, because of the large search areas we consider ($\sim$248
arcmin$^2$).  To illustrate how our search results help us constrain
the overall shape of the $UV$ LF, we show in Figure~\ref{fig:expcount}
the predicted counts for each of our search fields from our best-fit
$UV$ LF (i.e., $M_{UV}^*=-19.8$ mag, $\phi=0.0011$ Mpc$^{-3}$) in red
and compare it against the best-fit model where $M_{UV}^*$ shows no
evolution from $z\sim4$ (green lines: $M_{UV}^*=-21$, $\phi^* =
0.00012$ Mpc$^{-3}$).  The observed surface densities of dropouts are
shown with the solid histogram.  As is apparent from this figure, the
model with the bright value of $M_{UV}^*$ predicts too many bright
sources (e.g., ISAAC/MOIRCS panel) and too few faint sources (e.g.,
HUDF panel), whereas the model with a fainter value of $M_{UV}^*$
provides a better match to the observations.  The significance of this
result, however, is clearly only modest ($3\sigma$) at present, given
the limited statistics.  Better statistics would also be helpful for
verifying that our treatment of the selection effects in each search
field is reasonably accurate.

For our $z\sim9$ $J$-dropout selection, we cannot derive a set of
best-fit Schechter parameters due to the lack of any dropout
candidates in our sample.  Nonetheless, it does make sense for us to
set constraints on $M_{UV}^*$ and $\phi^*$ assuming no evolution in
the other Schechter parameters from $z\sim4$.  For example,
supposing there is no evolution in $M_{UV}^*$ (i.e., $M_{UV}^*=-21$),
we can set a $1\sigma$ upper limit of $<0.00006$ Mpc$^{-3}$
($\sim20\times$ lower than at $z\sim4$) on $\phi^*$.  Similarly,
assuming there is no evolution in $\phi^*$ (i.e., $\phi^*=0.0011$
Mpc$^{-3}$), we can set a $1\sigma$ lower limit of $-19.6$ AB mag (1.4
mag fainter than at $z\sim3$) on $M_{AB} ^{*}$.  The joint constraints
on the value of $\phi^*$ and $M_{UV}^*$ are shown in
Figure~\ref{fig:contourml} with the black lines indicating the 68\%
and 95\% confidence limits.

\begin{figure}
\epsscale{1.20}
\plotone{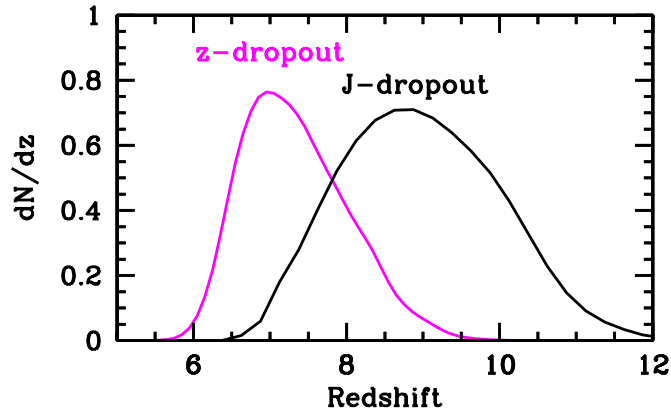}
\caption{The redshift distributions predicted for our $z$ and
$J$-dropout selections (shown as the magenta and black lines,
respectively) using our derived LFs and the simulation results
presented at the beginning of \S4.  The $z$-dropout predictions assume
$M_{UV,AB}=-19.8$ and $\alpha=-1.74$ (Table~\ref{tab:lfparm}) while
the $J$-dropout prediction assume $M_{UV,AB}=-19.1$ and $\alpha=-1.74$
(derived from the extrapolations derived in \S5.3).  The mean
redshifts expected for our z-dropout and J-dropout selections are 7.3
and 9.0, respectively.  These mean redshifts are somewhat lower than
that of previous selections we had considered (e.g., Bouwens et al.\
2004c; Bouwens et al.\ 2005; Bouwens \& Illingworth 2006) because of
our use of slightly more inclusive selection criteria in this
work.\label{fig:zdist}}
\end{figure}

In constraining the Schechter parameters at $z\sim7$ and $z\sim9$
from our $z$ and $J$-dropout searches, we assumed that the faint-end
slope $\alpha$ was equal to the same value we had previously
determined at $z\sim6$, i.e., $\alpha=-1.74$.  To see how these
constraints might depend upon the assumed faint-end slope $\alpha$, we
repeated the determinations assuming a faint-end slope $\alpha$ of
$-1.4$ and $-2.0$.  The results are presented in
Table~\ref{tab:lfparmalpha}, and it is clear that the derived values
for $M_{UV}^*$ and $\phi^*$ only show a mild dependence on the assumed
faint-end slope $\alpha$, particularly when compared to the
uncertainties on these two parameters from other error sources (e.g.,
from small number statistics).  We can therefore essentially ignore
the effect of the uncertain faint-end slope on our quoted constraints
for both $M_{UV}^*$ and $\phi^*$.

To put the LF fit results and dropout selections in context, it seems
helpful to use our best-fit LFs to calculate an expected redshift
distribution for our $z$ and $J$ dropout selections.  We can take
advantage of the same simulations described at the beginning of \S4 to
calculate these redshift distributions.  For concreteness,
$M_{UV,AB}^{*}=-19.1$ and $\alpha=-1.74$ are assumed for the $UV$ LF
at $z\sim9$ for our $J$-dropout selection (using the extrapolations
from \S5.3).  The redshift distributions are presented in
Figure~\ref{fig:zdist}.  The mean redshifts expected for our z-dropout
and J-dropout selections are 7.3 and 9.0, respectively.  The mean
redshifts estimated for both selections are smaller than what we had
considered in previous works (e.g., Bouwens et al.\ 2004c; Bouwens et
al.\ 2005; Bouwens \& Illingworth 2006) because of our adoption of a
slightly more inclusive dropout criterion in this work.  Note that
including the likely evolution across the redshift range of these
selections would likely lower the mean redshifts even more (Mu{\~n}oz
\& Loeb 2008).

\begin{figure}
\epsscale{1.20}
\plotone{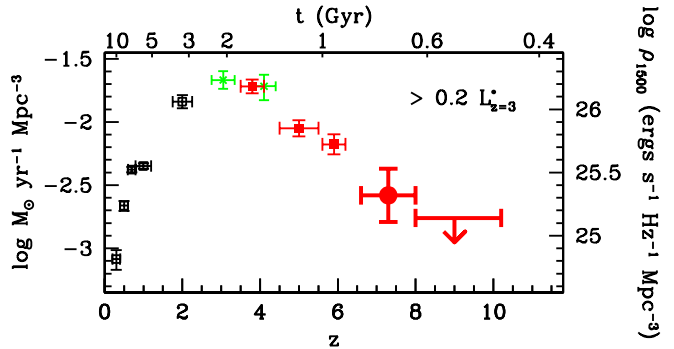}
\caption{The present constraints on the $UV$ luminosity density at
$z\gtrsim7$.  At $z\sim7$, this constraint is shown as a large solid
red circle while at $z\sim9$, it is shown as a $1\sigma$ upper limit
(\textit{red downward arrow}).  These determinations are integrated to
$0.2L_{z=3}^{*}$ to match the approximate faint limits on our
$z\gtrsim7$ galaxy searches.  Also shown are the determinations of
Schiminovich et al.\ (2005: \textit{open black squares}), Steidel et
al.\ (1999: \textit{green crosses}), and Bouwens et al.\ (2007:
\textit{solid red squares}) integrated to the same flux
limit.\label{fig:sfz}}
\end{figure}

\begin{figure}
\epsscale{1.20}
\plotone{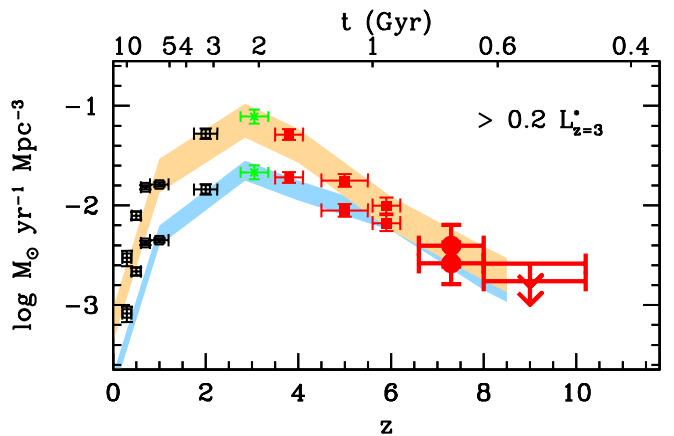}
\caption{Estimated star formation rate density as a function of
redshift (integrated down to 0.2 $L_{z=3}^{*}$ as in
Figure~\ref{fig:sfz}).  The lower set of points give the SFR density
without a correction for dust extinction, and the upper set of points
give the SFR density with such a correction.  This is also indicated
with the shaded blue and red regions, respectively, where the width of
these regions show the approximate uncertainties estimated by
Schiminovich et al.\ (2005).  At lower redshift ($z\lesssim3$), we
adopt the dust correction suggested by Schiminovich et al.\ (2005).
At $z\gtrsim6$, we adopt the dust correction obtained by Bouwens et
al.\ (2006: see also Stark et al.\ 2007a and Stanway et al.\ 2005) at
$z\sim6$ from the $UV$-continuum of $i$-dropouts and the Meurer et
al.\ (1999) IRX-$\beta$ prescription.  At $z\sim4$ and $z\sim5$, we
interpolate between the estimated dust extinctions at $z\sim3$ and
$z\sim6$.  The symbols are the same as in
Figure~\ref{fig:sfz}.\label{fig:dustsfz}}
\end{figure}

\begin{deluxetable}{lcccc}
\tablewidth{0pt}
\tabletypesize{\footnotesize}
\tablecaption{$UV$ Luminosity Densities and Star Formation Rate Densities.\tablenotemark{a}\label{tab:sfrdens}}
\tablehead{
\colhead{} & \colhead{} & \colhead{$\textrm{log}_{10} \mathcal{L}$} & \multicolumn{2}{c}{$\textrm{log}_{10}$ SFR density} \\
\colhead{} & \colhead{} & \colhead{(ergs s$^{-1}$} & \multicolumn{2}{c}{($M_{\odot}$ Mpc$^{-3}$ yr$^{-1}$)} \\
\colhead{Sample} & \colhead{$<z>$} & \colhead{Hz$^{-1}$ Mpc$^{-3}$)} & \colhead{Uncorrected} & \colhead{Corrected\tablenotemark{b}}}
\startdata
$z$ & 7.3 & 25.32$\pm$0.21 & $-2.58\pm0.21$ & $-2.40\pm0.21$ \\
$J$ & 9.0 & $<25.14$\tablenotemark{c} & $<-2.76$\tablenotemark{c} & $<-2.58$\tablenotemark{c} \\
\multicolumn{5}{c}{-----------------------------------------------------------------} \\
$B$ & 3.8 & 26.18$\pm$0.05 & $-1.72\pm$0.05 & $-1.29\pm0.05$ \\
$V$ & 5.0 & 25.85$\pm$0.06 & $-2.05\pm$0.06 & $-1.75\pm0.06$ \\
$i$ & 5.9 & 25.72$\pm$0.08 & $-2.18\pm$0.08 & $-2.00\pm0.08$ \\
\enddata
\tablenotetext{a}{Integrated down to 0.2 $L_{z=3}^{*}$.  Based upon LF parameters in Table~\ref{tab:lfparm} (see \S4.3).}

\tablenotetext{b}{The adopted dust-extinction corrections are 0.4 mag,
  0.6 mag, and 1.1 mag at $z\gtrsim6$, $z\sim5$, and $z\sim4$,
  respectively.  Our use of an evolving dust correction from
  $z\gtrsim6$ to $z\sim3$ is motivated by the apparent evolution in
  $UV$-continuum slope over this redshift range (e.g., Stanway et al.\ 2005;
  Bouwens et al.\ 2006) and the correlation of $UV$-continuum slope
  with dust extinction (e.g., Meurer et al.\ 1999; Reddy et al.\
  2006).}
\tablenotetext{c}{Upper limits here are $1\sigma$ (68\%
  confidence).}
\end{deluxetable}

\subsection{Luminosity / SFR density}

We can use the present LF results to refine our constraints of the
rest-frame $UV$ luminosity densities at $z\gtrsim7$.  Previously, we
set constraints on these quantities based upon a sample of four
$z\sim7$ $z$-dropout candidates found over the GOODS fields (Bouwens
\& Illingworth 2006) and a sample of three possible $J$-dropout
candidates over the deep NICMOS parallels to the HUDF (Bouwens et al.\
2005).

Here we are able to make modest but significant improvements in our
measurement of the $UV$ luminosity density at $z\sim7-10$.  We make
these estimates based upon our best-fit LFs at $z\sim7$ and
$z\sim9$.  We shall consider integrations to a limiting luminosity of
0.2 $L_{z=3}^{*}$ because this corresponds to the approximate flux
limit for our $z$ and $J$-dropout searches (i.e., $H_{160,AB}\sim28$).
We present these luminosity densities in Figure~\ref{fig:sfz} and
Table~\ref{tab:sfrdens}.  As is conventional, we also present the
equivalent star formation rate on this same figure using the Madau et
al.\ (1998) and adopting a Salpeter IMF.  To show how this situation
changes when a plausible account of dust is included, we plot the
equivalent dust-corrected and uncorrected SFR densities in
Figure~\ref{fig:dustsfz}.  We employ the dust corrections adopted by
Bouwens et al.\ (2007), which are 1.4 mag, 1.1 mag, 0.6 mag, and 0.4
mag at $z\lesssim3$, $z\sim4$, $z\sim5$, and $z\gtrsim6$,
respectively.  Our use of an evolving dust correction from
$z\gtrsim6$ to $z\sim3$ is motivated by the apparent evolution in
the $UV$-continuum slope over this redshift range (e.g., Stanway et
al.\ 2005; Bouwens et al.\ 2006) and the correlation of $UV$-continuum
slope with dust extinction (e.g., Meurer et al.\ 1999; Reddy et al.\
2006).

\section{Discussion}

\subsection{How does the volume density of $UV$-bright galaxies evolve?}

Over the past few years, there has been some controversy regarding how
the UV LF evolves at high redshift.  While some studies have argued
that the evolution primarily occurs at the bright end (i.e., Dickinson
et al.\ 2004; Shimasaku et al.\ 2005; Ouchi et al.\ 2004a; Bouwens et
al.\ 2006; Yoshida et al.\ 2006; Bouwens et al.\ 2007), there have
been other efforts which have argued that the evolution occurs
primarily at the faint end (i.e., Iwata et al.\ 2003; Sawicki \&
Thompson 2006; Iwata et al.\ 2007).  In this work, we found additional
evidence to support the fact that the most rapid evolution occurs at
the bright end of the $UV$ LF and that this evolution can be
approximately described by a change in the characteristic luminosity
$M_{UV}^*$ of the UV LF.

A significant part of the debate regarding the form the evolution of
the $UV$ LF takes at high redshift is centered on what happens at the
bright end of the LF (i.e., $M_{UV,AB}\lesssim-20.5$).  Does it evolve
or not?  Iwata et al.\ (2007) and Sawicki \& Thompson (2006) argue
that there is little evidence for substantial evolution from $z\sim5$
to $z\sim3$ from their work.  However, a quick analysis of results in
the literature indicate that such evolution is certainly very strong,
particularly if the baseline is extended out to even higher redshifts
($z\gtrsim6$).  At $z\sim6$, for example, the bright end of the $UV$
LF (i.e., $M_{UV,AB}<-20.5$) has been reported to be deficient by
factors of $\sim6-11$ relative to $z\sim3-4$ (Stanway et al.\ 2003,
2004; Shimasaku et al.\ 2005; Bouwens et al.\ 2006).  At $z\gtrsim7$,
this deficit is even larger, as we can see by comparing the number of
$UV$ bright galaxies (i.e., $M_{UV,AB}<-20.5$) in our dropout
selections with that expected from $z\sim4$ (Bouwens et al.\ 2007)
assuming no-evolution.  The comparisons are presented in
Table~\ref{tab:noevol}.  At $z\sim7$, the volume density of $UV$
bright ($M_{UV,AB}<-20.5$) galaxies appears to be
$18_{-11}^{+32}\times$ smaller than at $z\sim4$ (see also Bouwens \&
Illingworth 2006; Mannucci et al.\ 2007; Stanway et al.\ 2008), and at
$z\sim9$, the volume density of $UV$ bright galaxies is at least
16$\times$ smaller (68\% confidence).

\begin{deluxetable}{cccc}
\tablecolumns{4} 
\tablecaption{Comparison of the Observed Numbers of Dropouts with
No-evolution Extrapolations from
$z\sim4$\tablenotemark{a}\label{tab:noevol}}

\tablehead{ \colhead{} & \multicolumn{2}{c}{Number} & \colhead{Evolutionary} \\
\colhead{Sample} & \colhead{Observed} & \colhead{$z\sim4$ Prediction\tablenotemark{a}} & \colhead{Factor\tablenotemark{b}}}
\startdata 
\multicolumn{3}{c}{Bright ($M_{UV,AB} < -20.5$)} \\
$z$-dropout & 2 & 35.9 & $18_{-11}^{+32}$ \\ 
$J$-dropout & 0 & 17.9 & $>16$\tablenotemark{c} \\
\multicolumn{3}{c}{Faint ($M_{UV,AB} > -20.5$)} \\ 
$z$-dropout & 6 & 24.7 & $4_{-2}^{+3}$ \\ 
$J$-dropout & 0 & 8.0 & $>7$\tablenotemark{c} \\ 
\enddata
\tablenotetext{a}{Extrapolations assume no evolution in the $UV$ LF
from $z\sim4$ (Bouwens et al.\ 2007) and were calculated using
Eq.~\ref{eq:numcountg}.  The sizes and $UV$ colors of $UV$-bright
galaxies at $z\gtrsim7$ assumed for these estimates are the same as
those given in Appendix B.}
\tablenotetext{b}{Ratio of volume density of galaxies at $z\sim4$
(Bouwens et al.\ 2007) in some luminosity range to that in our higher
redshift dropout samples.}
\tablenotetext{c}{Lower limits here are $1\sigma$ (68\% confidence).}
\end{deluxetable}

The only analysis to find mild evolution in the volume density of $UV$
bright galaxies from $z\sim7-10$ to $z\sim3-4$ is that of Richard et
al.\ (2006) and involves a search for $z\geq6$ galaxies behind lensing
clusters.  However, as we argue in Appendix C, it seems likely that a
substantial fraction of the $UV$-bright galaxies in the Richard et
al.\ (2006) $z\sim6-10$ selection are spurious and therefore the
bright end of their $UV$ LFs is much too high.  The evidence for this
is rather striking: not only are none of the $z\geq6$ candidates in
the Richard et al.\ (2006) selection detected ($<$$2\sigma$) in
significantly deeper ($\sim$1-2 mag) NICMOS+IRAC data (11 of their
$z\geq6$ candidates), but also their reported prevalence is
$>$$10\times$ higher than what we measure in searches for similar
candidates behind lensing clusters (Bouwens et al.\ 2008; see also
Appendix C).

Considering the substantial evolution found from $z\gtrsim7$ to
$z\sim4$ in the volume density of $UV$ bright galaxies, it is
certainly relevant to note that all three previous studies (i.e.,
Iwata et al.\ 2003; Sawicki \& Thompson 2006; Iwata et al.\ 2007)
which argued for less evolution at the \textit{bright} end of the LF
than at the \textit{faint} end did not extend to redshifts much higher
than 4.  In fact, the very mild evolution (or no-evolution) at the
bright end of LF claimed by the above studies seems to be highly
consistent with other analyses in this redshift range (e.g., Steidel
et al.\ 1999; Reddy et al.\ 2007) which in general find very little
evolution from $z\sim4$ to $z\sim2$.  Even for the Iwata et al.\
(2007) LF at $z\sim4.7$ (which is fairly similar to the $UV$ LFs at
$z\sim2-4$ at bright luminosities), one may not expect much evolution
at the bright end.  After all, the mean redshift of their bright
selection is $z\sim4.5$, which is not much higher than
$z\sim4$.\footnote{The mean redshift for bright galaxies in the Iwata
et al.\ (2007) selection is likely to be somewhat lower in redshift
than for the fainter ones.  This is because (1) the Iwata et al.\
(2007) selection window extends to somewhat lower redshift for the
redder star-forming galaxy population than it does for the bluer
star-forming galaxy population and (2) brighter star-forming galaxies
at $z\sim4-6$ are observed to have much redder UV colors in general
than the fainter star-forming galaxies at these redshifts (e.g.,
Meurer et al.\ 1999; Iwata et al.\ 2007; R.J. Bouwens et al.\ 2008, in
prep).  As a result, the mean redshift for bright galaxies in the
Iwata et al.\ (2007) selection will be lower than for the faint
galaxies, and therefore the bright end of the Iwata et al.\ (2007) LF
should show less evolution than the faint end.}

In summary, it would appear that the most reasonable conclusions to
draw from these studies are that the bright end of the $UV$ LF shows
substantial evolution at very high redshifts ($z\gtrsim4$) and then
mild evolution at somewhat later times (i.e., from $z\sim4$ to
$z\sim2$).

\subsection{Completeness of current samples in $UV$-continuum slope}

The selection criteria we use to select star-forming galaxies at
$z\gtrsim7$ (and at $z\lesssim6$) differ somewhat in the efficiency
with which they select galaxies that have red $UV$-continuum slopes
$\beta$, i.e., $\beta\gtrsim0.0$.\footnote{For our $z$-dropout
selection, for example, we would only be moderately effective at
selecting galaxies with $UV$-continuum slopes $\beta$ redder than
$\beta=0$ (Figure~\ref{fig:zjjh}).  For our $J$-dropout selection, on
the other hand, there is no limit on how red a star-forming galaxy can
be in terms of its $UV$-continuum slope, except as set by our
restrictions on the IRAC and MIPS colors of sources (i.e.,
$3.6\mu-5.8\mu < 1$, $3.6\mu - 24\mu < 2$: see \S3.2) which require
$\beta\lesssim0$.  For the Bouwens et al.\ (2007) $z\sim4$
$B$-dropout, $z\sim5$ $V$-dropout, and $z\sim6$ $i$-dropout
selections, we are only effective in selecting star-forming galaxies,
if their $UV$-continuum slopes $\beta$ are bluer than $\beta=0.5$ (see
Figure 8 of Bouwens et al.\ 2007).}  This would potentially have an
effect on our determinations of the $UV$ LFs if such red star-forming
galaxies existed in large numbers.  Fortunately, for these dropout
selections, it appears that they do not.  At $z\sim4$, for example,
both Lyman-break and Balmer-break selections almost entirely consist
of galaxies with $\beta$'s bluer than $-1.0$ (\S4.1 and Figure 8 of
Bouwens et al.\ 2007; Brammer \& van Dokkum 2007).  At $z\sim5-6$, the
galaxy population appears to be bluer yet, with mean $\beta$'s of
$\sim-2$ (e.g., Stanway et al.\ 2005; Bouwens et al.\ 2006; Yan et
al.\ 2005; Lehnert \& Bremer 2003; Hathi et al.\ 2008), and there is
little evidence from the present selections that the galaxy population
is any less blue at $z\gtrsim7$.  Indeed, the observed
$J_{110}-H_{160}$ colors are consistent with a star-forming population
with little dust and $10^8$ years of constant star formation (where
$\beta \sim -2$: see Figure~\ref{fig:comp}).

\begin{deluxetable}{cccccc}
\tablewidth{0pt}
\tabletypesize{\footnotesize}
\tablecaption{Best-fit schechter parameters for the $UV$ LFs at
$z\sim7$ and $z\sim9$ assuming different mean $UV$-continuum slopes
$\beta$'s to compute the selection volumes for our dropout
samples (see \S5.2).\tablenotemark{a}\label{tab:betadep}}
\tablehead{\colhead{Dropout} & \colhead{} & \colhead{Mean} & \colhead{} & \colhead{$\phi^*$ $(10^{-3}$} & \colhead{} \\
\colhead{Sample} & \colhead{$<z>$} & \colhead{$\beta$} &
\colhead{$M_{UV} ^{*}$\tablenotemark{b}} & \colhead{Mpc$^{-3}$)}}
\startdata
$z$ & 7.3 & $-2.0$ & $-19.8\pm0.4$ & $1.1_{-0.7}^{+1.6}$\\
$z$ & 7.3 & $-2.5$ & $-19.8\pm0.4$ & $1.3_{-0.8}^{+1.7}$\tablenotemark{d}\\
$z$ & 7.3 & $-1.5$ & $-19.8\pm0.4$ & $1.3_{-0.8}^{+1.7}$\tablenotemark{d}\\
$J$ & 9.0 & $-2.0$ & $>-19.6$\tablenotemark{c} & $(1.1)$\tablenotemark{e}\\
$J$ & 9.0 & $-2.5$ & $>-19.7$\tablenotemark{c} & $(1.1)$\tablenotemark{e}\\
$J$ & 9.0 & $-1.5$ & $>-19.5$\tablenotemark{c} & $(1.1)$\tablenotemark{e}\\
\enddata
\tablenotetext{a}{All LFs assume a faint-end slope $\alpha$ equal to $-1.74$.}
\tablenotetext{b}{See remarks in Table~\ref{tab:lfparm}.}
\tablenotetext{c}{Lower limits here are $1\sigma$ (68\% confidence).}
\tablenotetext{d}{Both bluer and redder $UV$-continuum slopes can
result in slightly lower estimated selection volumes for star-forming
galaxies at $z\sim7$.}
\tablenotetext{e}{Fixed in our fits.}
\end{deluxetable}

Of course, even if our dropout selections are largely complete in
their selection of dropouts over a wide range in $UV$-continuum slope,
it is possible that the $UV$ LF we derive may depend somewhat on the
distribution of $UV$ continuum slopes we are using to model the
selection volumes for our high-redshift dropout samples.  We would not
expect the distribution we are assuming to be too inaccurate, as we
are able to use this distribution to reproduce the observed
$J_{110}-H_{160}$ colors of our $z$-dropout sample (Appendix B: see
Figure~\ref{fig:comp}).  Nonetheless, it is worthwhile examining the
sensitivity of our results to small changes in the distribution of
$UV$-continuum slopes $\beta$.  Bouwens et al.\ (2007) examined the
effect that different assumptions about these slopes would have on the
$UV$ LF at $z\sim4-6$ and found changes of $\lesssim10$\% in the fit
parameters for shifts of 0.7 in the mean $UV$-continuum slope $\beta$.
Repeating our determinations of the $UV$ LF at $z\sim7$ and $z\sim9$
assuming a mean $UV$-continuum slope of $-2.5$, we find differences of
$\lesssim20$\% in the constraints on $M^*$ and $\phi^*$
(Table~\ref{tab:betadep}).  Similarly small differences were found
relative to previous fits assuming a mean $UV$-continuum slope $\beta$
of $-1.5$ (Table~\ref{tab:betadep}).  This suggests that our LF fit
results are relatively robust with respect to uncertainties about the
distribution in $UV$-continuum slopes $\beta$.

\begin{figure}
\epsscale{1.20}
\plotone{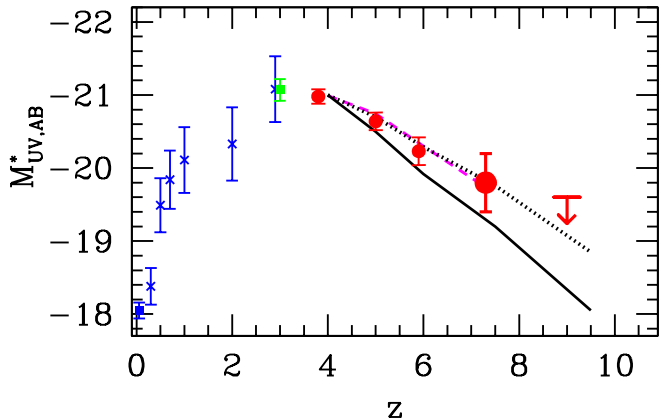}
\caption{Determinations of the characteristic luminosity $M^*$ in the
rest-frame $UV$ as a function of redshift.  Shown are our present
determination at $z\sim7$ (large red circle) and $1\sigma$ upper limit
at $z\sim9$ (red downward arrow).  Also included are the
determinations at $z\sim4$, $z\sim5$, and $z\sim6$ from Bouwens et
al.\ (2007: \textit{red circles}), at $z\sim3$ from Steidel et al.\
(1999: \textit{green cross}), and at $z\lesssim3$ from Arnouts et al.\
(2005: \textit{blue crosses}), and at $z\sim0$ from Wyder et al.\
(2005: \textit{blue square}).  The expected evolution in $M^*$
predicted from the momentum-driven wind ``vzw'' model of Oppenheimer
\& \dave\,(2006) is plotted as the dashed magenta line while the
evolution expected as a result of a build-up in the mass function is
shown.  The solid black line shows the evolution predicted from the
halo mass function (Sheth \& Tormen 1999) assuming a mass-to-light
ratio which is constant while the dotted black line shows this
evolution assuming a mass-to-light ratio which varies as $(1+z)^{-1}$
(which is somewhat less than the $(1+z)^{-1.5}$ scaling of the
mass-to-light ratio used in Wyithe \& Loeb 2006 or Stark et al.\
2007c: see \S5.4).  The predicted evolution in $M_{UV}^*$ is derived
from the models by examining the changes in the LF (or mass function)
at a fixed volume density $\sim10^{-2.5}$ Mpc$^{-3}$.  The above
predictions differ from those given in Bouwens et al.\ (2007) due to
our use of a more current input matter power spectrum (Spergel et al.\
2007).  The observed evolution in $M_{UV}^*$ appears to be in
reasonable agreement with what one might expect from the evolution of
the halo mass function, assuming an evolution in the mass-to-light
ratio for halos of $\sim(1+z)^{-1}$.\label{fig:abmagz}}
\end{figure}

\subsection{Evolution of the LF}

In this work, we derived constraints on the $UV$ LF at $z\sim7$ and
$z\sim9$ using both a stepwise and Schechter parametrization.  For
our Schechter parametrization, we estimated that the value of
$M_{UV}^*$ for the $UV$ LF at $z\sim7$ was equal to $-19.8\pm0.4$
mag and $\phi^*=0.0011_{-0.0007}^{+0.0017}$ Mpc$^{-3}$.  The value of
$M_{UV}^*$ we derived at $z\sim7$ is $0.4\pm0.4$ mag fainter than we
found at $z\sim6$.  This is consistent with the trend previously
discovered over the redshift interval $z\sim4-6$ where the value of
$M_{UV}^*$ was fainter the earlier back in time we probed (see also
Yoshida et al.\ 2006).  We have already remarked that the value of
$\phi^*$ is quite consistent with the values found at $z\sim4$,
$z\sim5$, and $z\sim6$ by Bouwens et al.\ (2007: see also Yoshida et
al.\ 2006).  We also attempted to constrain the Schechter parameters
at $z\sim9$ using the results of our $J$-dropout search.  Assuming
that we can model the changes in the $UV$ LF from $z\sim9$ using a
Schechter function and that these changes can parametrized by an
evolution in $M_{UV}^*$ (with $\phi^*$ and $\alpha$ fixed, as is
suggested by our constraints on the UV LF from $z\sim4$ to
$z\sim7$), we found that $M_{UV,z=10} ^{*} > -19.6$ mag (68\%
confidence).  Plotting the constraints on $M_{UV}^*$ versus redshift
(Figure~\ref{fig:abmagz}), it seems clear that the observations are
consistent with a substantial brightening of $M_{UV}^*$ at early
cosmic times.  The brightening of $M_{UV}^*$ is $>1.4$ mag from
$z\sim9$ to $z\sim4$.  This, of course, is in contrast to the
behavior of $M_{UV}^*$ at $z\lesssim4$ where $M_{UV}^*$ reaches a
maximum brightness at $z\sim2-4$ and then fades with cosmic time
(e.g., Gabasch et al.\ 2004; Arnouts et al.\ 2005).

We can put together the constraints on the $UV$ LF from $z\sim7-10$ to
$z\sim4$ (i.e., Figure~\ref{fig:contourml} from this work and Figure 3
from Bouwens et al.\ 2007) to derive approximate expressions for
$M_{UV}^*$, $\phi^*$, and $\alpha$ as a function of redshift (at
$z\gtrsim4$).  Assuming that the dependence of these parameters on
redshift can be parametrized in a linear way and moreover that the the
$UV$ LF can be represented by a Schechter function at $z\gtrsim7$ (see
discussion in \S5.5), we derive the following relationships for $M_{UV}^*$,
$\phi^*$, and $\alpha$ for $z\sim4-7$:
\begin{eqnarray*}
M_{UV} ^{*} =& (-21.02\pm0.09) + (0.36\pm0.08) (z - 3.8)\\
\phi^* =& (1.16\pm0.20) 10^{(0.024\pm0.065)(z-3.8)}10^{-3} \textrm{Mpc}^{-3}\\
\alpha =& (-1.74\pm0.05) + (0.04\pm0.05)(z-3.8)
\label{eq:empfit}
\end{eqnarray*}
These equations provide us with a convenient way of extrapolating our
observational results back to early times (i.e., $z>7$).  They
also provide us with a very quantitative way of assessing the
significance of the evolutionary trends identified in $M_{UV}^*$,
$\phi^*$, and $\alpha$.  In particular, these equations suggest that
the observed evolution in $M_{UV}^*$ is significant at the $4.5\sigma$
level over the interval $z\sim9$ to $z\sim4$, while the evolution in
$\phi^*$ and $\alpha$ is still not significant at all (i.e.,
$<1\sigma$).

\subsection{Relation to the Halo Mass Function}

The rapid brightening of galaxies within the first two billion years
of the universe is not particularly surprising since it is during this
time period that $L^*$ galaxy halos (typically $10^{10}$ to $10^{12}$
$M_{\odot}$: e.g., Cooray 2005; Lee et al.\ 2006; Ouchi et al.\ 2004b;
Cooray \& Ouchi 2006) are largely assembled.  The volume density of
$\sim10^{11}$ $M_{\odot}$ halos is expected to increase 1000 fold from
$z\sim10$ to $z\sim3-4$ (see, e.g., Figure 2 of Springel et al.\
2005).  Larger increases are expected at $\gtrsim10^{11}$ $M_{\odot}$,
and smaller increases are expected at $\lesssim10^{11}$ $M_{\odot}$.
Thus, if the efficiency of star formation in galaxies does not
decrease substantially as a function of cosmic time, we would expect
substantial increases in the volume density of luminous star-forming
galaxies from $z\sim9$ to $z\sim4$.  We would also expect the maximum
typical luminosity of star-forming galaxies (and thus $M_{UV}^*$) to
increase over this time period (due to the greater evolution expected
for more massive galaxies).  This is exactly what our LFs suggest
(Figure~\ref{fig:abmagz}).

In light of these expectations, it is relevant to try to connect the
evolution we observe in the rest-frame $UV$ LF to the anticipated
changes in the halo mass function.  Not only does this provide us with
a simple test of our theory for structure formation, but it also
allows us to place modest constraints on the efficiency with which
halos in the early universe produce $UV$ photons.

In our previous work (\S5.2 of Bouwens et al.\ 2007), we decided to
parametrize the evolution in the $UV$ LF and that expected for the
halo mass function in terms of a single variable for simplicity.
Motivated by our observational findings (Bouwens et al.\ 2006; Bouwens
et al.\ 2007), we took this parameter to be the characteristic
luminosity for our $UV$ LF and some characteristic mass for the halo
mass function.  The characteristic mass was taken to equal the mass
where the mass function had a comoving volume density of $10^{-2.5}$
Mpc$^{-3}$ to (approximately) match the volume density of $\phi^*$
galaxies.  We ignored questions about a possible evolution in the
shape of the $UV$ LF or mass function.\footnote{We note that we find
very similar changes in the so-called characteristic mass here if we
make the comparisons in terms of the average (or even median) mass of
halos at the high-mass end of the mass function ($\lesssim10^{-2.5}$
Mpc$^{-3}$).}  Making different assumptions about how the
mass-to-light ratio (halo mass to $UV$ light) of halos vary as a
function of redshift, we developed predictions for how $M_{UV}^*$
might be expected to change with redshift -- assuming a matter power
spectrum with $\sigma_8 = 0.9$, spectral index $n_s=1$,
$\Omega_M=0.3$, $\Omega_{\Lambda}=0.7$, and $H_0 =
70\,\textrm{km/s/Mpc}$.  We found that the observations seemed to
prefer a scenario where there was little evolution in the
mass-to-light ratio of halos at early times.

Given the similarity of the present observational results to those
previously considered by Bouwens et al.\ (2007), one might suspect
that the above conclusions would remain unchanged.  However, after
exploring the sensitivity of these conclusions to different inputs, we
found that they could depend significantly upon assumptions made in
calculating the halo mass function.  In particular, we found that for
the matter power spectrum preferred by the WMAP three year data and
the SDSS large-scale structure constraints ($\sigma_8 = 0.772$,
$n_s=0.948$), the evolution we predicted for the characteristic mass
was $\sim$40\% larger than what we predicted in our previous work
using $\sigma_8=0.9$ and $n_s=1$.  This translated into a $\sim$40\%
larger predicted brightening of $M_{UV}^{*}$ with cosmic time than in
the previous predictions, and so as a result our models with no
evolution in the mass-to-light ratio (halo mass to UV light) were no
longer in good agreement with the observations (see solid line in
Figure~\ref{fig:abmagz}).

We therefore explored different scalings of the mass-to-light ratio of
halos to try to regain good agreement with the observations.  We found
reasonable agreement, adopting a halo mass-to-light ratio that varied
as $(1+z)^{-1}$ (see dotted line in Figure~\ref{fig:abmagz}).  Such a
scaling would mean that star formation at early times is much more
efficient than at later times and, in fact, does not scale that
differently from what one would expect based upon the dynamical time
scales (which evolves as $\sim(1+z)^{-1.5}$: e.g., Wyithe \& Loeb
2006; Stark et al.\ 2007c).  Of course, it seems prudent to remember
that these conclusions are likely highly uncertain.  They are clearly
somewhat sensitive to the input matter power spectrum that one adopts,
and it seems likely that there are other uncertainties that will have
an effect (e.g., the issues discussed in \S5.5).

\subsection{Shape of the $UV$ LFs at $z\gtrsim4$}

Of course, the $UV$ LFs we have derived at $z\gtrsim4$ contain a lot
more information than is present in a single parameter.  They also
contain substantial information on the overall shape of the $UV$ LF.
Perhaps the most notable feature in the shape of the observed LFs at
$z\sim4$, $z\sim5$, and $z\sim6$ is the presence of a distinct
``knee,'' where the LF transitions from a power-law-like behavior
towards fainter magnitudes to an exponential cut-off at the bright
end.  At lower redshifts, this exponential cut-off at the bright end
of the LFs is well-known and has been hypothesized to occur as a
result of a variety of different factors.  Classically, this cut-off
has been explained as a result of the inefficiency with which gas can
cool in halos above $10^{12}$ $M_{\odot}$ (e.g., Binney 1977; Rees \&
Ostriker 1977; Silk 1977) and settle out of hydrostatic equilibrium.
This explanation has recently been updated and recast in terms of a
discussion of hot and cold flows (i.e., Birnboim \& Dekel 2003; Dekel
\& Birnboim 2004).  The exponential cut-off at the bright end of the
LF has also been explained as a result of AGN feedback (e.g., Binney
2004; Scannapieco \& Oh 2004; Croton et al.\ 2006; Granato et al.\
2004), which above a certain mass scale is thought to cut off the flow
of cold gas to a galaxy and therefore stop star formation.  One final
factor which may be important for imparting the sharp cut-off at the
bright end of the LF, particularly at $UV$ wavelengths, is dust
obscuration, which becomes substantial (factors $\gtrsim10$) for the
most luminous systems (e.g., Wang \& Heckman 1996).  As a result, the
$UV$ luminosity of star-forming galaxies does not tend to reach above
some maximum value (see discussion in e.g., Adelberger \& Steidel
2000; Martin el al.\ 2005).

\begin{figure}
\epsscale{1.20}
\plotone{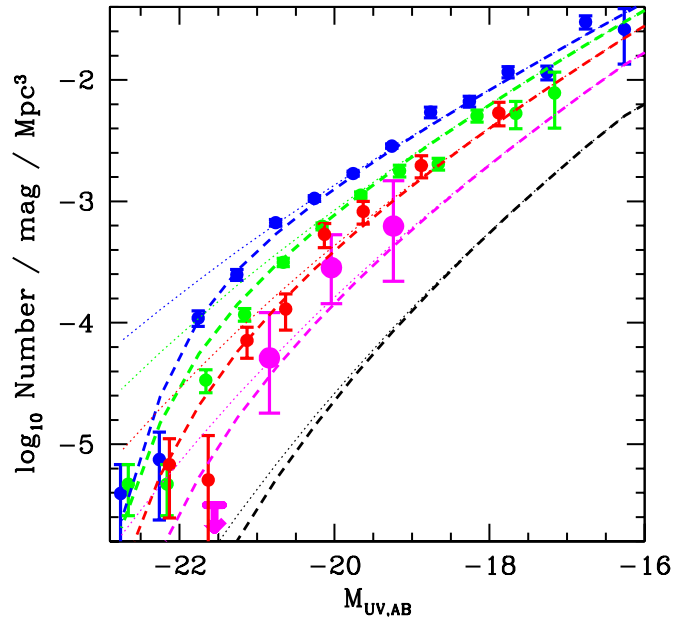}
\caption{Rest-frame $UV$ LFs predicted at $z\sim4$, $z\sim5$,
$z\sim6$, $z\sim7$, and $z\sim9$ (\textit{dashed blue, green, red,
magenta, and blacks lines, respectively}) using a CLF formalism (see
\S5.5) compared with those from our observations.  The stepwise LFs
found in the observations are presented in the same way as in
Figure~\ref{fig:lf}.  The CLF kernel used to convert mass into light
was chosen to fit the LF at $z\sim4$.  The predicted LFs using a
kernel with no cut-off at the bright end of the mass function is shown
using the dotted line (i.e., assuming $L_c (M) =
(2.51\times10^{22}\,\textrm{W}\, \textrm{Hz}^{-1}) (M/m_c)^{1.24}$).
At $z\gtrsim4$, we adopt the same CLF kernels as at $z\sim4$ to
calculate the UV LF from the halo mass function, but modified to
include the $\sim(1+z)^{-1}$ evolution in the mass-to-light ratio
inferred by comparing the $UV$ LFs with the expected halo mass
function (see Figure~\ref{fig:abmagz}).  The LFs predicted using these
assumptions do not show an extremely prominent ``knee'' so
characteristic of ``typical'' Schechter functions at $z\geq6$.
\label{fig:clf}}
\end{figure}

At very high redshifts, where the typical halo masses are smaller,
none of the above mechanisms would seem to work very well.  The
typical masses are below those where AGN feedback might be expected to
be important (e.g., Scannapieco \& Oh 2004; Dekel \& Birnboim 2004) or
where gas cooling is inefficient (i.e. $<10^{12}\,M_{\odot}$: Dekel \&
Birnboim 2004).\footnote{Nonetheless, it has been argued (e.g., Dekel
\& Birnboim 2004) that at $z>2$ gas cooling will occur very
efficiently in some regions of high mass ($\gtrsim10^{12}\,M_{\odot}$)
halos through ``cold flows.''}  The expected star-formation rates at
these redshifts would also tend to be much lower and therefore dust
obscuration would not likely be such an important factor either (given
the correlation between these two quantities: e.g., Wang \& Heckman
1996).  As such, it would seem that there are no obvious mechanisms to
impose an abrupt cut-off at the bright end of the $UV$ LF and
therefore we might expect that the UV LFs at very early times
($z\gtrsim6$) to lack the ``knee'' so characteristic of the Schechter
parametrization and look much more like the mass function.  In fact,
the $UV$ LFs at $z\gtrsim4$ derived from a number of recent
theoretical simulations (e.g., Figure 7 of Finlator et al.\ 2006;
Figure 10 of Oesch et al.\ 2007) have exactly those characteristics.
Those predicted LFs bear a much closer resemblance to power laws than
they do to the traditional Schechter function.  The observation of a
significant knee--or cut-off--at the bright end of the observed LFs
out to $z\sim6$ may mean that there is some additional physics
relevant to the formation of galaxies that is not fully understood.

{\it Conditional Luminosity Functions:} To illustrate this issue
quantitatively, we adopt a simplified version of the conditional
luminosity function (CLF) formalism laid out in Cooray \& Ouchi (2006:
see also Yang et al.\ 2003; Cooray \& Milosavljevi{\'c} 2005).  The
CLF formalism provides us with a way of converting some halo mass
function to a LF.  Typically there is a kernel which when convolved
with the mass function gives the LF, i.e.,
\begin{displaymath}
\phi(L) = \int_M \phi(L|M) \frac{dN}{dM} dM
\end{displaymath}
where
\begin{eqnarray*}
\phi(L|M) &= \frac{1}{\sqrt{2\pi}(\log_{e} 10)\sigma L} \times \\
           & ~ ~ \exp \left\{ - \frac{\log_{10} [L/L_c (M)]^2}{2\sigma^2} \right\}
\end{eqnarray*}
where $\frac{dN}{dM}$ is the Sheth-Tormen (1999) mass function, where
$\log_{e} 10\approx2.303$, where $L_c (M)$ gives the $UV$ luminosity
of the central galaxy in some halo of mass $M$ and where the parameter
$\sigma$ expresses the dispersion in the relationship between the halo
mass and the $UV$ light of the central galaxy.  Note that we have
ignored the contribution from satellite galaxies to the luminosity
function in the above equation since they appear to constitute
$\lesssim10$\% of the galaxies over a wide-range in luminosity (see,
e.g., Cooray \& Ouchi 2006).  One aspect about the CLF formalism is
that it is largely independent of much of the normal phenomenology
associated with galaxy formation and therefore it is very clean.
There is no attempt to explicitly model the large number of physical
processes likely to be important in the formation and evolution of
galaxies (which remain a challenge for most semi-analytic
prescriptions, e.g., Somerville \& Primack 1999; Cole et al.\ 2000).

Making use of the above formalism, we experimented with a variety of
different parameterizations to find a function $L_c (M)$ and
parameters $\sigma$ and $m_c$ which reproduced the Bouwens et al.\
(2007) $UV$ LF at $z\sim4$.  We found that $L_c (M) =
(2.51\times10^{22}\,\textrm{W}\, \textrm{Hz}^{-1})
\frac{(M/m_c)^{1.24}}{(1+(M/m_c))}$, $\sigma=0.16$, and $m_c =
1.2\times10^{12}\,M_{\odot}$ worked reasonably well (see the blue
dashed line in Figure~\ref{fig:clf}).  $2.51\times10^{22}\,
\textrm{W}\,\textrm{Hz}^{-1}$ is equivalent to $-21.91$ AB mag.  We
adopted the best-fit values of $\sigma_8$, $n_s$, $\Omega_{m}$,
$\Omega_{\Lambda}$, $\Omega_b$, and $h$ to the WMAP three year + SDSS
data (Spergel et al.\ 2007).  Our choice of $m_c$ set the mass scale
where the exponential cut-off in the luminosity function takes effect.
At the high mass end of the mass function, our equation for $L_c$
asymptotes towards a $(M/m_c)^{0.2}$ behavior.  The physical
motivation for this cut-off at the bright-end of the mass function
would presumably be the same as at low redshift, where star formation
is not as efficient for the most massive halos because of inefficient
gas cooling or AGN feedback.  To show the effect the high mass cutoff
has on the model LF, we also show this LF without such a cut-off (see
the dotted lines in Figure~\ref{fig:clf}).  At the low-mass end, our
equation for $L_c$ aymptotes towards a $(M/m_c)^{1.24}$ behavior.
This allowed us to account for the fact that the faint-end slope of
the $UV$ LF is slightly shallower than that of the mass function
(i.e., $-1.7$ vs. $-2.0$).  The chosen parameters yield similar,
albeit slightly lower, halo masses to those inferred from clustering
analyses of star-forming galaxies at $z\sim4$ (e.g., Ouchi et al.\
2004b; Lee et al.\ 2006).

The question now is whether the above relationship between $UV$ light
and halo mass (i.e., the kernel $\phi(L|M)$) can reproduce the
rest-frame $UV$ LF at $z\gtrsim4$.  Given the reasonable agreement
between the observed evolution in $M_{UV}^*$ and that predicted from
the halo mass function assuming $\sim(1+z)^{-1}$ in the mass-to-light
ratio, one might expect that the above kernel may work if we multiply
the $L_c (M)$ expression above to include the factor
$(\frac{1+z}{1+3.8})$.  With this change, $L_c (M)$ equals
$(2.51\times10^{22}\,\textrm{W}\, \textrm{Hz}^{-1})
\frac{(M/m_c)^{1.24}}{(1+(M/m_c))} (\frac{1+z}{1+3.8})$.  The UV LFs
we predict at $z\sim5$, $z\sim6$, $z\sim7$, and $z\sim9$ are
presented in Figure~\ref{fig:clf} as the green, red, magenta, and
black dashed lines, respectively.  These LFs appear to differ somewhat
in overall shape from what is found in our observations, predicting
somewhat fewer galaxies at fainter luminosities than are observed,
particularly near the ``knee'' of the LF (i.e., at $\sim-19.5$ AB
mag).  Most notably, the sharp jump in the volume densities of sources
observed between the bright end of the LF (i.e., $\sim-22$ AB mag) and
the ``knee'' of the LF (i.e., $\sim-20$ AB mag) is not well matched by
the predicted LFs at $z\gtrsim6$.

This effectively illustrates what we were describing above.  To obtain
a LF with a prominent ``knee,'' we require some physical mechanism
which will truncate the LF above some luminosity (or mass).  Since
this ``knee'' of the LF would seem to shift to much fainter
luminosities at early cosmic times (assuming that the evolution of the
$UV$ LF can be effectively parametrized in terms of $M_{UV}^*$ to
$z\sim9$ as suggested by Figure~\ref{fig:abmagz} or
Eq.~\ref{eq:empfit}), this mechanism needs to become important at
progressively lower luminosities (and also lower masses), as we move
back in cosmic time.  This is in contrast to the assumptions we use in
the formalism above, where we have assumed that the mass cut-off $m_c$
is independent of cosmic time.  It is not clear that any of the
mechanisms proposed to produce a cut-off at the bright end of the LF
(e.g., AGN feedback, dust extinction, inefficient gas cooling at high
masses) will extend to very low luminosities (or masses).  The
observations may therefore present us with a bit of puzzle.
Alternatively, it is also possible given the observational
uncertainties that the $UV$ LFs at $z\gtrsim7$ will increasingly look
like the halo mass function and less like a Schechter function with a
prominent ``knee.''

\section{Summary}

We have taken advantage of all the deep near-IR and optical data that
are currently public over and around the two GOODS fields (including
the HUDF) to search for $z\gtrsim7$ $z$ and $J$-dropout candidates.
In total, we make use of $\sim23$ arcmin$^2$ of deep NICMOS data and
$\sim248$ arcmin$^2$ of deep ACS+ISAAC/MOIRCS data.  Over these areas,
we identify 8 $z$-dropout candidates, but no plausible $J$-dropout
candidates.  By exploring all reasonable sources of contamination
(T-dwarfs, high-redshift supernovae, photometric scatter, spurious
sources) for our modest $z$-dropout sample, we argue that the majority
of sources in our selection ($\gtrsim88$\%) are bona-fide star-forming
galaxies at $z\sim7$.  Using the size-redshift trends, $UV$ colors,
and spatial profiles of star-forming galaxies at $z\sim4-6$, we make
plausible estimates of the selection volume for identifying $z\sim7$
$z$-dropouts and $z\sim9$ $J$-dropouts.  We then use these samples and
selection volumes to derive the LF at $z\sim7$ and set constraints on
the LF at $z\sim9$.  We consider both stepwise and Schechter
parameterizations for the $UV$ LF.

Our primary conclusions are:

\textit{(1) UV LF at $z\sim7$:} The rest-frame $UV$ LF shows
significant evolution from $z\sim7$ to $z\sim6$ (99\% confidence).
Our best-fit $\phi^*$ and $M_{UV}^*$ for the $UV$ LF at $z\sim7$ is
$0.0011_{-0.0007}^{+0.0017}$ Mpc$^{-3}$ and $-19.8\pm0.4$ mag,
respectively, adopting the same value for the faint-end slope $\alpha$
($-1.74$) that we previously found at $z\sim6$.  This value of
$M_{UV}^*$ is nearly $1.2\pm0.4$ mag fainter than the value of
$M_{UV}^*$ that has previously been found at $z\sim4$ and $0.4\pm0.4$
mag fainter than the value found at $z\sim6$.  By contrast, the value
of $\phi^*$ that we determined at $z\sim7$ is not significantly
different from those values determined at slightly later times
($z\sim4-6$: Bouwens et al.\ 2007).  Similar to several previous
studies (Bouwens et al.\ 2006; Yoshida et al.\ 2006; Bouwens et al.\
2007), this suggests that the evolution of the $UV$ LF at $z\gtrsim4$
can largely be described by an evolution in the characteristic
luminosity $M_{UV}^*$, with no significant changes required in the
other variables.

\textit{(2) Constraints on the UV LF at $z\sim9$:} Since we did not
find any credible $J$-dropout candidates in our search fields, we were
not able to derive best-fit parameters for the $UV$ LF at $z\sim9$ and
could only set constraints on $\phi^*$ and $M_{UV}^*$.  To interpret
the implications of our $J$-dropout search results for the $UV$ LF at
$z\sim9$, we assumed that the value of $\phi^*$ and $\alpha$ there
were roughly equal to the values (i.e., 0.0011 Mpc$^{-3}$ and $-1.74$)
found at $z\sim4-8$.  We then derived lower limits on the value of
$M_{UV}^*$ by using our $J$-dropout search results and the effective
selection volumes we calculated for $z\sim9$ galaxies using numerous
Monte-Carlo simulations.  We found that $M_{UV} ^* > -19.6$ mag at
68\% confidence.  These search constraints are inconsistent with no
evolution from $z\sim9$ to $z\sim7$ at 80\% confidence.

\textit{(3) Evolution in the volume density of $UV$-bright galaxies:}
Comparing the no-evolution predictions of the Bouwens et al.\ (2007)
$z\sim4$ UV LF at bright magnitudes with those observed in the ISAAC,
MOIRCS, or NICMOS data (see Table~\ref{tab:noevol}: see \S5.1), we
estimated that the bright end ($M_{UV,AB}<-20.5$) of the $UV$ LF is at
least $18_{-10}^{+32}\times$ lower at $z\sim7$ than it is at
$z\sim4$.  Making a similar comparison between the number of
$J$-dropouts predicted assuming no evolution from $z\sim4$ and the
zero candidates found in our NICMOS search fields, we estimated that
the bright end of the LF ($M_{UV,AB}<-20.5$) is at least $16\times$
lower at $z\sim9$ than it is at $z\sim4$ (68\% confidence).  This
provides strong evidence against evolutionary models with little
evolution at the bright end of the $UV$ LF.

\textit{(4) Empirical fit to the UV LF from $z\sim9$ to $z\sim4$:}
Putting together our constraints on the $UV$ LF from $z\sim9$ to
$z\sim4$, we derived fitting formula to describe the evolution of
$M_{UV}^*$, $\phi^*$, and $\alpha$ as a function of redshift.  The
results are $M_{UV} ^{*} = (-21.02\pm0.09) + (0.36\pm0.08) (z - 3.8)$,
$\phi^* = (1.16\pm0.20) 10^{(0.024\pm0.065)(z-3.8)}10^{-3}
\textrm{Mpc}^{-3}$, and $\alpha = (-1.74\pm0.05) +
(0.04\pm0.05)(z-3.8)$.  These formulae imply that the evolution in
$M_{UV}^*$ is significant at the $4.5\sigma$ level.  The evolution in
$\phi^*$ and $\alpha$ is not yet significant.

\textit{(5) Relationship between the evolution of the $UV$ LF at
$z\gtrsim4$ and that expected for the halo mass function:} The
evolution we observe in the characteristic luminosity $M_{UV}^*$
versus redshift matches the evolution expected for the halo mass
function, assuming that the mass-to-light ratio (halo mass to UV
light) of halos evolves as $\sim(1+z)^{-1}$ with cosmic time.  This
finding is somewhat different from what Bouwens et al.\ (2007)
inferred about the mass-to-light ratios of halos based upon very
similar observational constraints.  The reason for the change is our
use of more current constraints on the matter power spectrum (i.e.,
$\sigma_8 = 0.772$, spectral index $n_s=0.948$: Spergel et al.\ 2007)
rather than the $\sigma_8=0.9$, $n_s=1$ power spectrum we were
assuming in Bouwens et al.\ (2007).  Effectively, this illustrates how
sensitive galaxy evolution can be at high-redshift to the input matter
power spectrum.

\textit{(6) Shape of the $UV$ LF at $z\gtrsim4$:} At redshifts as
early as $z\sim6$, the $UV$ LF is remarkably well-represented by a
Schechter function (Bouwens et al.\ 2007).  The position of the
``knee'' of the LF is known to $\lesssim0.2$ mag at $z\lesssim6$
(Bouwens et al.\ 2007) and would seem to be known to $\lesssim0.4$ mag
at $z\sim7$.  However, it is unclear why the $UV$ LF would show a
prominent knee at such early times.  Most of the physical mechanisms
thought to create this knee in the LF at lower redshifts (e.g., AGN
feedback or a post-virialization cooling condition) would not seem to
be effective at the lower masses relevant at these times.\vskip0.2cm

The constraints we have on the $UV$ LF at $z\gtrsim7$ are already
quite notable, especially given the significant challenges that exist
to the detection of $z\gtrsim7$ galaxies with current observational
facilities.  Putting together all the data now available, we can
plausibly determine the volume density of $UV$-bright galaxies at
$z\sim7$ to within a factor of 2 and set meaningful constraints on
the volume density of similarly luminous galaxies at $z\sim9$.
Comparing the present search results with those obtained at lower
redshift ($z\sim4-6$), we observe a rather substantial increase in the
volume density of most luminous $UV$-bright galaxies, from redshifts
$z\gtrsim7$ to $z\sim4$ -- suggesting that such luminous and likely
prodigious star-forming systems only start to become common relatively
late in cosmic time (i.e., 1.5 Gyr after the Big Bang).  These
galaxies are presumably built up through the merging and coalescence
of lower mass galaxies.

Over the next year, we expect our constraints on the bright end of
$UV$ LF at $z\gtrsim7$ to improve substantially as a result of both
ground and space-based imaging programs.  From the ground, these
improvements will come as a result of the deep wide-area imaging being
obtained over the GOODS fields with wide-area near-IR imagers such as
MOIRCS on Subaru telescope and HAWK-I on VLT-Yepes.  From space, these
improvements will come from two large-to-medium-sized HST NICMOS
programs (GO-11082 and GO-11144 are surveying $\sim60$ arcmin$^2$ to a
depth of $\sim26.5$ AB).  Of course, for truly significant advances in
the size and depth of current samples, we will need to wait until late
2008 when WFC3 with its powerful IR channel will be installed on the
Hubble Space Telescope.

\acknowledgements

We would like to thank Louis Bergeron, Susan Kassin, Dan Magee,
Massimo Stiavelli, and Rodger Thompson for their assistance in the
reduction of NICMOS data which has been essential to find and
characterize faint star-forming galaxies at $z\gtrsim7$.  We
acknowledge stimulating conversations with Peter Capak, Michele
Cirasuolo, Asantha Cooray, Romeel \dave, James Dunlop, Richard Ellis,
Kristian Finlator, Brad Holden, Cedric Lacey, Ross McLure, Pascal
Oesch, Masami Ouchi, Naveen Reddy, Piero Rosati, and Daniel Stark.  We
are grateful to Kristian Finlator for his continuing helpfulness in
computing rest-frame $UV$ LFs at $z\gtrsim4$ from his group's
theoretical models.  Ivo Labb{\'e} provided us with a very deep
38-hour reduction of the VLT and Magellan $K_s$ band imaging data over
the HUDF.  We thank Pascal Oesch and Li Xin for a careful read of our
manuscript.  We acknowledge support from NASA grants HST-GO09803.05-A
and NAG5-7697.

\appendix

\section{A.  Large Scale Structure Uncertainties}

\begin{figure}
\epsscale{1.1}
\plotone{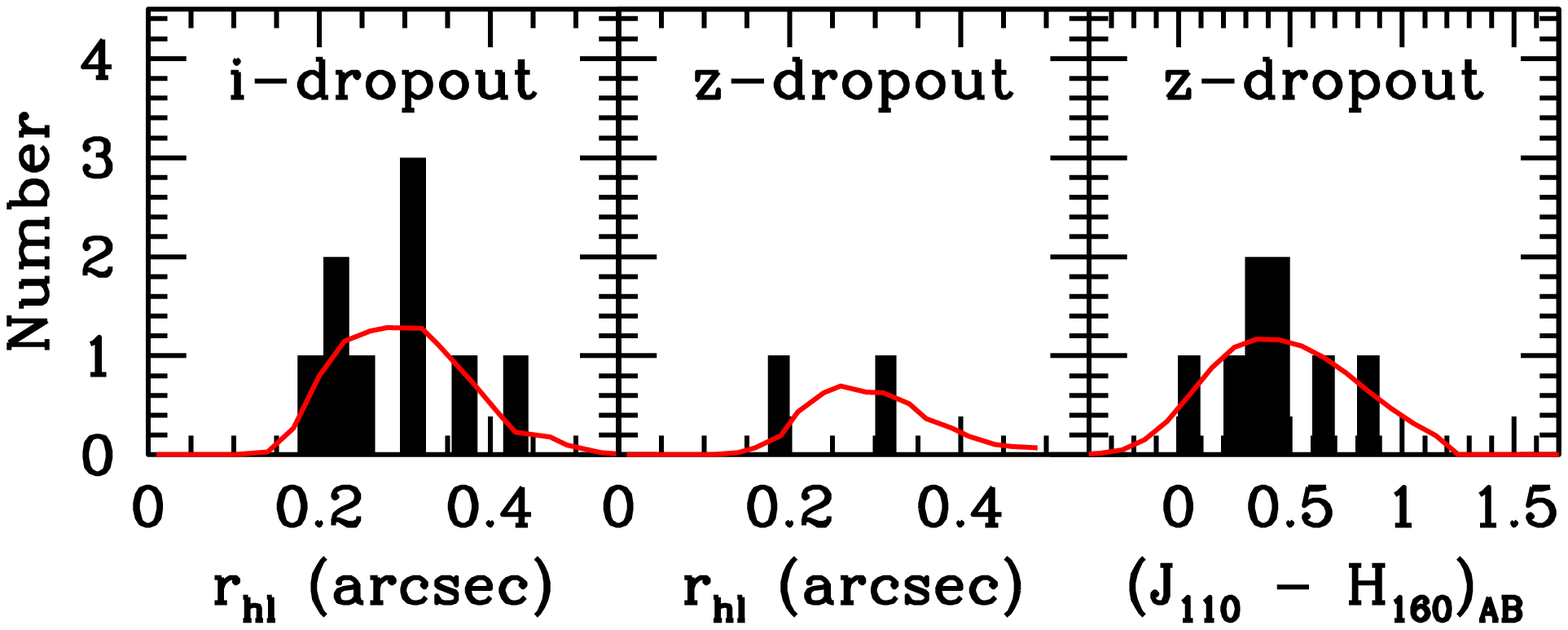}
\caption{Comparison between the observed properties of the $i$ and $z$
dropouts (black histogram) and that expected based on our model for
star-forming galaxies at $z\gtrsim6$ (red lines: see \S4 and Appendix
B for a description of this model).  The leftmost panels show
comparisons between the observed and expected half-light radii
($r_{hl})$ while the rightmost panel shows comparisons between the
observed and expected $UV$ colors.  Only galaxies brightward of
$H_{160,AB}=26.4$ and $H_{160,AB}=27.0$ in our HUDF and NICPAR
selections are included in the distribution of half-light radii.  This
leaves us with some leverage in examining the size (surface
brightness) distribution.  Otherwise, selection effects would tend to
dominate, excluding all the larger sources from these comparisons and
we would be left with only those sources whose sizes were very similar
to that of the PSF.  For the comparisons we make involving $UV$
colors, all search fields are considered.  In general, good agreement
is found between the observed and expected properties, providing us
with a reasonable amount of confidence in the model we are using to
calculate the effective selection volumes for star-forming galaxies at
$z\gtrsim7$.\label{fig:comp}}
\end{figure}

In this work, we derived our best-fit LF at $z\sim7$ by comparing
the observed surface density of galaxies with what we would expect
based on a model LF and then maximizing the likelihood of the model
LF.  We assumed Poissonian statistics in computing the likelihood of
various model LFs (\S4).  Our decision to adopt a Poissonian model in
computing these likelihoods rather utilizing a more conventional
approach like SWML was to take advantage of the substantial amount of
deep, wide-area near-IR imaging (e.g., the ISAAC or MOIRCS data) we
have to constrain the $UV$ LF at $z\sim7$.  Since most of these
imaging data contain no $z\sim7$ candidates, we would not be able to
use that data for improving our determination of the $UV$ LF at
$z\sim7$, had we chosen to use the conventional SWML or STY79
techniques.  However, modelling our observational results using
Poissonian statistics (and assuming no large-scale structure
variations), we are able to take advantage of this information to
constrain the shape of the $UV$ LF.

Of course, in modelling the observational results using this approach,
we are inherently assuming that the volume density of high-redshift
star-forming galaxies is statistically homogeneous across the sky.  In
reality, however, the distribution of galaxies is modestly clustered,
and therefore the mean volume density of galaxies will show
significant variations as a function of position.  These variations in
the mean density occur as a result of large-scale structure and will
have an effect on the best-fit $UV$ LF we derive at $z\sim7$,
scattering it away from the cosmic average.

To estimate the effect that such variations will have on our fit
results, we ran a set of Monte-Carlo simulations.  We varied the
surface density of model dropouts in each of our search fields by
$\sim30$\% RMS (see \S4.1) and then recomputed the $UV$ LF in the same
way as in \S 4.2.  We found $1\sigma$ variations of $\pm0.1$ mag in
the values of $M_{UV}^*$ and $\pm$30\% in the values of $\phi^*$ at
$z\sim7$.  Adding these uncertainties in quadrature with those
deriving from Poissonian uncertainties (\S 4.2), the total uncertainty
in $\phi^*$ at $z\sim7$ is 0.4 dex and $M_{UV}^*$ is $\pm0.4$ mag.

The effect of large-scale structure variations on the constraints we
place on the $UV$ LF at $z\sim9$ is much smaller in general.  Given
that we have no sources in our $J$-dropout sample, the Poissonian
uncertainties ($\pm\sim100$\%) dominate over the expected
field-to-field variations, which are $\pm\sim20$\% (see \S4.2 or
Bouwens et al.\ 2005).

\section{B.  Plausiblity of our model for estimating the selection volumes of dropout samples at $z\gtrsim7$}

Besides small number statistics, perhaps the biggest uncertainty in
the present determination of the rest-frame $UV$ LF at $z\gtrsim7$ is
in our estimates of the selection volumes for the two dropout samples.
We derived these estimates using the following model for the sizes,
morphologies, and $UV$ colors of galaxies at $z\gtrsim7$.  We assumed
that star-forming galaxies at $z\gtrsim7$ have similar pixel-by-pixel
profiles to the $z\sim4$ $B$-dropouts in the HUDF, but with physical
sizes scaled by a factor of $(1+z)^{-1.1}$ (e.g., at $z\sim7$ galaxies
of a given luminosity have approximately half the physical size of
star-forming galaxies at $z\sim4$ and therefore four times the surface
brightness).  The $UV$-continuum slopes $\beta$ were taken to have a
mean of $-2.0$ and $1\sigma$ scatter of 0.5.  Both assumptions were
chosen to match the trends found in the observations from $z\sim6$ to
$z\sim2$ for the sizes (e.g., Bouwens et al.\ 2006; Bouwens et al.\
2004a; Ferguson et al.\ 2004) and colors (Stanway et al.\ 2005;
Bouwens et al.\ 2006).

Of course, just because the present model is a reasonable
extrapolation of the trends (in sizes and colors) found at lower
redshift does not mean it will accurately represent star-forming
galaxies at $z\gtrsim7$.  Therefore, it makes sense for us to compare
this model with the limited data on $z\gtrsim6$ galaxies where they
exist.  We begin by comparing the observed $J-H$ colors of
$z$-dropouts in our sample with that predicted using our best-fit LF
($\phi\sim0.0011$ Mpc$^{-3}$, $M^{*}_{1900,AB}\sim-19.8$ mag,
$\alpha\sim-1.74$) and the model we just described above (see also
\S4) for calculating the selection volumes.  We include the comparison
in the rightmost panel of Figure~\ref{fig:comp}.

Similarly, we can compare the half-light radii of $z$-dropouts in our
sample with that predicted from our models.  To ensure that the size
distribution we derive for $z$-dropouts is not dominated by surface
brightness selection effects (which only allows for the selection of
point-like sources near the magnitude limit), we only consider the
sizes of $z$-dropouts that are $\gtrsim0.7$ mag brighter than the
magnitude limit.  Unfortunately, this only leaves us with $\sim2$
galaxies for making this comparison.  The result is shown in the
middle panel of Figure~\ref{fig:comp}.  The model predictions appear
to be in reasonable agreement with the observations.  As one
additional check on the size-redshift scaling we are using to model
the sizes (surface brightnesses) of $z\sim7$ galaxies, we also
include a comparison of our model predictions with the half-light
radii of $z\sim6$ $i$-dropouts in our search fields (leftmost panel of
Figure~\ref{fig:comp}).

\section{C.  Comparison with the Richard et al.\ (2006) results}

In the present work (e.g., \S5.1 or Table~\ref{tab:noevol}), we found
a significantly smaller (i.e., $\sim1/18$) volume density of galaxies
at the bright end of the $UV$ LF at $z\gtrsim7$ than has been inferred
at $z\sim4$ (e.g., Steidel et al.\ 1999; Reddy et al.\ 2007; Bouwens
et al.\ 2007).  This is in significant contrast with the volume
densities of $z\gtrsim6$ galaxies inferred by Richard et al.\ (2006)
from searches for $z$ and $J$ dropouts in $\sim12$ arcmin$^2$ of deep
($\sim24.5$ AB mag at $5\sigma$) near-IR data behind two low-redshift
lensing clusters Abell 1835 and AC114.  Richard et al.\ (2006) found
13 first and second category $z$ and $J$ dropouts in their search that
they consider plausible $z>6$ galaxies.  Modelling the lensing
amplication, contamination, and overall incompleteness of their
search, Richard et al.\ (2006) estimated the $UV$ LFs at $z\sim6-8$
and $z\sim8-10$.  In significant contrast to our results, the volume
density of bright galaxies in the Richard et al.\ (2006) LFs were
found to be very similar to what has been found at $z\sim3-4$ (i.e.,
Steidel et al.\ 1999; Bouwens et al.\ 2007).

What is the reason for these substantial differences?  While one
possible explanation could be ``cosmic variance'' (i.e.,
field-to-field variations), simple estimates for the variance one
expects in the volume density of high-redshift sources behind a
typical cluster is of order $\sim$50\% RMS, and therefore much smaller
than the factor of $\gtrsim10$ differences seen here.  These ``cosmic
variance'' estimates assume a pencil beam geometry, a redshift window
of width $\Delta z = 1$, an angular diameter of $\sim30$ arcmin$^2$,
and a bias of $\sim7$, which is appropriate for galaxies at $z\sim7$
with a volume density of $\sim10^{-3.5}$ Mpc$^{-3}$ (see also
Somerville et al.\ 2004; Trenti \& Stiavelli 2008).

Instead a much more probable explanation for the discrepancy seen here
is simply that a substantial fraction of the Richard et al. (2006)
$z\gtrsim6$ candidates are spurious in nature.  After all, none of
their best candidates were detected at significance levels much higher
than $2.5\sigma$, making each of the candidates in their sample open
to question.  Clearly, given the area of empty sky within their 12
arcmin$^2$ search area, it would not be difficult to imagine a small
number of these empty areas of sky showing large enough positive
fluctuations to satisfy their detection thresholds just by chance!  To
their credit, Richard et al.\ (2006) recognized the significance of
this concern and attempted to model this effect.  After extensive
Monte-Carlo simulations, they argued that the number of dropouts found
in their samples significantly exceeded that expected from the noise.
Unfortunately, such techniques can be highly sensitive to even the
smallest errors in the noise model, and therefore it is not entirely
clear how much weight we should give to the simulations Richard et
al.\ (2006) have run to control for this process.

A much more straightforward and robust way of estimating the spurious
fraction is to look at their candidates in deeper near-IR data (as
Bremer et al.\ 2004 used to evaluate the reality of the Pell{\'o} et
al.\ 2004 $z\sim10$ candidate).  Such data are available for two deep
$\sim0.7$ arcmin$^2$ NIC3 $H_{160}$ exposures over each of their two
clusters (AC114 and Abell1835).  These NICMOS pointings reach to
$\sim26.8$ AB mag (5 $\sigma$), $\sim2$ mag deeper than the ISAAC data
used by Richard et al.\ (2006) in their searches.  These pointings
provide very deep $H_{160}$-band coverage for 2 first and second
category $z\gtrsim6$ candidates reported by Richard et al.\ (2006) and
5 first, second, and third category candidates.  The first and second
category candidates of Richard et al.\ (2006) were the $z\gtrsim6$
candidates to which they ascribed the highest confidence while the
third category candidates were ascribed slightly lower confidence.  Of
the 5 such candidates with very deep $H_{160}$-band coverage, none
show up at $>2\sigma$ significance, strongly arguing that a
substantial fraction of their candidates do not correspond to real
sources.  Taken by itself, it implies that $\gtrsim80$\% of the
Richard et al.\ (2006) first, second, and third category $z\gtrsim6$
candidates are spurious (68\% confidence).  If we only consider first
and second category candidates (the highest confidence candidates)
with deep NICMOS data, the NICMOS non-detections imply that
$\gtrsim56$\% of the Richard et al.\ (2006) candidates are spurious
(68\% confidence).

We also do not find it particularly encouraging that none of the
dropout candidates in the Richard et al.\ (2006) sample are detected
in the deep IRAC $3.6\mu$ imaging which exist around these clusters
(Schaerer et al.\ 2007).  There are six sources in the Richard et al.\
(2006) sample which are sufficiently isolated from the neighbors for
these upper limits to have particular significance.  Given the
reported $H$-band magnitudes of the Richard et al.\ (2006) $z\gtrsim7$
candidates (ranging from 24.6 to 25.2 AB mag), we would suspect it
should be easy to detect these sources in the very deep IRAC data
available (reaching to 25.8 AB mag at $2\sigma$), especially given
that typical $(z_{850}-3.6\mu m)_{AB}$ colors for other $z\sim6-8$
candidates range from $-0.5$ mag to $1.5$ mag (Yan et al.\ 2006;
Labb{\'e} et al.\ 2006; Eyles et al.\ 2007; Bradley et al.\ 2008).
While Schaerer et al.\ (2007) suggest that the IRAC non-detections
argue for a very young population, we do not think the candidates'
reality is well enough established to draw any conclusions about the
properties of the purported sources.  Rather we find it worrying that
none of the available NICMOS or IRAC data over these clusters have
provided us with evidence that any of the Richard et al.\ (2006)
candidates are real!

Finally, we note the overall prevalence of $z\gtrsim6$ candidates
found by Richard et al.\ (2006) behind AC114 and Abell 1835 is at
least $\gtrsim10\times$ higher than found in searches around other
lensing clusters with much deeper, higher quality NICMOS data.
Bouwens et al.\ (2008) have quantified this by examining the deep
($\gtrsim26.5$ AB mag at $5\sigma$) $J$ and $H$-band imaging data
around 6 galaxy clusters (CL0024, MS1358, Abell 2218, Abell 2219,
Abell 2390, Abell 2667) and the deep ($\gtrsim26$ AB mag at $5\sigma$)
$J$-band imaging around three other galaxy clusters (Abell 1689, Abell
1703, 1E 0657-56).  In total, the deep near-IR imaging data around
these clusters cover $\gtrsim20$ arcmin$^2$ in total, almost double
what was considered in Richard et al.\ (2006) and extending some
$\gtrsim1-2$ mag deeper.  Only one $z\gtrsim7$ candidate (i.e.,
A1689-zD1: Bradley et al.\ 2008) and one $z\sim6.5$ candidate (Kneib
et al.\ 2004) is found over these fields to the same magnitude limit
(i.e., 25.5 AB mag) as is considered in the Richard et al.\ (2006)
search.  This is to be compared with 13 first and second category
$z\gtrsim6$ candidates found by Richard et al.\ (2006) in a search
over just half the area.  Since it seems unlikely that high-redshift
dropouts are $\gtrsim10\times$ more common behind the two particular
clusters considered by Richard et al.\ (2006) than the 9 other
clusters with deep NICMOS data, this again suggests that the majority
(i.e., $\gtrsim90$\%) of the $z\gtrsim6$ candidates found in the
Richard et al.\ (2006) search are simply spurious sources.

\section{D.  $z\gtrsim7$ Dropouts in previous samples not in current selections}

There were a small number of candidates in our previous samples (3
$z$-dropout candidates from Bouwens et al. 2004c and 3 $J$-dropout
candidates from Bouwens et al. 2005) which did not make it into our
current selections.  In this section, we describe why those particular
candidates are not included in our current selections.

Let us first consider the three $z$-dropout candidates from the HUDF
(Bouwens et al.\ 2004c) which do not make it into our current samples.
As has already been discussed in Bouwens \& Illingworth (2006) and
Labb{\'e} et al. (2006), two of those candidates (UDF-818-886 and
UDF-491-880) were the direct result of the ``Mr. Staypuft'' anomaly
(Skinner et al.\ 1998), which produces faint electronic ghosts 128
pixels from bright stars.  Since the ``Mr. Staypuft'' anomaly is now
carefully modelled and explicitly removed in our ``nicred.py''
reductions (Magee et al.\ 2007), this anomaly should not pose a
problem for our current selections.  The third candidate (UDF-825-950)
may still be a high-redshift star-forming galaxy, but did not match
the S/N thresholds used for the current selection.

Now we consider the three $J$-dropout candidates reported by Bouwens
et al.\ (2005).  What became of these three candidates?  Two of the
three candidates (UDFNICPAR1-04151142 and UDFNICPAR2-09593048) show
detections in the ultra-deep $V$-band imaging recently taken over the
NICMOS parallels to the HUDF and therefore are clearly not $z\sim9-10$
galaxies.  The other candidate (UDFNICPAR1-05761077) is not detected
at high enough significance in our current reductions of the NICMOS
parallels to the HUDF and appears to have been spurious.  Spurious
sources are a particular concern over the NICMOS parallels to the HUDF
given the minimal amounts of dithering used in acquiring the data.
Therefore, substantial care has been taken to accurately characterize
the read-out properties of individual pixels and correct for any
pixel-by-pixel detector (or reduction) signatures (see \S 2.1).

The loss of these three $z\sim9-10$ $J$-dropout candidates from our
samples should not be a significant cause for concern.  In the Bouwens
et al.\ (2005) analysis, the emphasis was on setting an \textit{upper
limit} on the volume density of $UV$-bright $z\sim9-10$ galaxies and
therefore we tried to err on the side of \textit{including} candidates
in our samples if there was any uncertainty about their nature.


\begin{thebibliography}{} 
\bibitem[Adelberger \& Steidel(2000)]{2000ApJ...544..218A} Adelberger, 
K.~L.~\& Steidel, C.~C.\ 2000, \apj, 544, 218
\bibitem[Arn]{2005arn} Arnouts, S., et al.\  2005, \apjl, 619, L43 
\bibitem[Baba et al.(2002)]{2002ASPC..281..298B} Baba, H., et al.\ 2002, 
Astronomical Data Analysis Software and Systems XI, 281, 298 
\bibitem[Beckwith et al.(2006)]{2006AJ....132.1729B} Beckwith, S.~V.~W., et 
al.\ 2006, \aj, 132, 1729
\bibitem[Bertin and Arnouts (1996)]{1996A&AS..117..393B} Bertin, E.\ and 
Arnouts, S.\ 1996, \aaps, 117, 39
\bibitem[Binney(1977)]{1977ApJ...215..483B} Binney, J.\ 1977, \apj, 215, 
483 
\bibitem[Binney(2004)]{2004MNRAS.347.1093B} Binney, J.\ 2004, \mnras, 347, 
1093
\bibitem[Birnboim \& Dekel(2003)]{2003MNRAS.345..349B} Birnboim, Y., \& 
Dekel, A.\ 2003, \mnras, 345, 349 
\bibitem[Bouwens, Broadhurst and Silk (1998)]{1998ApJ...506..557B} Bouwens,
R., Broadhurst, T.\ and Silk, J.\ 1998a, \apj, 506, 557
\bibitem[Bouwens, Broadhurst and Silk (1998)]{1998ApJ...506..579B}
Bouwens, R., Broadhurst, T.\ and Silk, J.\ 1998b, \apj, 506, 579.
\bibitem[Bouwens, Broadhurst, \&
Illingworth(2003)]{2003ApJ...593..640B} Bouwens, R., Broadhurst, T.,
\& Illingworth, G.\ 2003, \apj, 593, 640
\bibitem[Bouwens et al.(2004)]{2004ApJ...606L..25B} Bouwens, R.~J., et al.\ 
2004a, \apjl, 606, L25
\bibitem[Bouwens et al.\ 2004]{b2004b} Bouwens, R.~J., Illingworth,
G.D., Blakeslee, J.P., Broadhurst, T.J., \& Franx, M.  2004b, \apjl,
611, L1
\bibitem[Bouwens et al.\ 2004]{b2004c} Bouwens, R.~J., et al.\  2004c, \apjl,
616, L79
\bibitem[Bouwens et al.\ 2005]{b2005} Bouwens, R.~J., Illingworth,
G.D., Thompson, R.I., \& Franx, M.  2005, \apj, 624, L5
\bibitem[Bouwens \& Illingworth(2006)]{2006Natur.443..189B} Bouwens, R.~J., 
\& Illingworth, G.~D.\ 2006, \nat, 443, 189.
\bibitem[Bouwens et al. 2006]{Bouwens} Bouwens, R.J., Illingworth,
G.D., Blakeslee, J.P., \& Franx, M.  2006, \apj, 653, 53 
\bibitem[Bouwens et al.(2007)]{2007ApJ...670..928B} Bouwens, R.~J., 
Illingworth, G.~D., Franx, M., \& Ford, H.\ 2007, \apj, 670, 928
\bibitem[Bouwens et al.(2008)]{2008ApJ...670..928B} Bouwens, R.~J., et
al.\ 2008, \apj, submitted, arXiv:0805.0593
\bibitem[Bradley]{Bradley} Bradley, L.D., et al. 2008, \apj, 678, 647
\bibitem[Brammer \& van Dokkum(2007)]{2007ApJ...654L.107B} Brammer, G.~B., 
\& van Dokkum, P.~G.\ 2007, \apjl, 654, L107 
\bibitem[Bremer et al.(2004)]{2004ApJ...615L...1B} Bremer, M.~N., Jensen, 
J.~B., Lehnert, M.~D., Schreiber, N.~M.~F., \& Douglas, L.\ 2004, \apjl, 
615, L1
\bibitem[Bruzual \& Charlot(2003)]{2003MNRAS.344.1000B} Bruzual, G., \& 
Charlot, S.\ 2003, \mnras, 344, 1000
\bibitem[Burgasser et al.(2006)]{2006ApJ...637.1067B} Burgasser, A.~J., 
Geballe, T.~R., Leggett, S.~K., Kirkpatrick, J.~D., \& Golimowski, D.~A.\ 
2006, \apj, 637, 1067
\bibitem[Coe et al.(2006)]{2006AJ....132..926C} Coe, D., Ben{\'{\i}}tez, 
N., S{\'a}nchez, S.~F., Jee, M., Bouwens, R., \& Ford, H.\ 2006, \aj, 132, 
926 
\bibitem[Cole et al.(2000)]{2000MNRAS.319..168C} Cole, S., Lacey, C.~G., 
Baugh, C.~M., \& Frenk, C.~S.\ 2000, \mnras, 319, 168
\bibitem[Coleman, Wu, \& Weedman(1980)]{1980ApJS...43..393C} Coleman, G.\ 
D., Wu, C.\ -., \& Weedman, D.\ W.\ 1980, \apjs, 43, 393
\bibitem[Cooray(2005)]{2005MNRAS.364..303C} Cooray, A.\ 2005, \mnras, 364, 
303
\bibitem[Cooray \& Milosavljevi{\'c}(2005)]{2005ApJ...627L..89C} Cooray, 
A., \& Milosavljevi{\'c}, M.\ 2005, \apjl, 627, L89
\bibitem[Cooray \& Ouchi(2006)]{2006MNRAS.369.1869C} Cooray, A., \& Ouchi, 
M.\ 2006, \mnras, 369, 1869 
\bibitem[Croton et al.(2006)]{2006MNRAS.365...11C} Croton, D.~J., et al.\ 
2006, \mnras, 365, 11
\bibitem[dejong]{dejong} de Jong, R.S., et al.\ 2006, The 2005 HST
Calibration Workshop, 121.
\bibitem[Dekel \& Birnboim(2006)]{2006MNRAS.368....2D} Dekel, A., \& 
Birnboim, Y.\ 2006, \mnras, 368, 2 
\bibitem[Dickinson(1998)]{1998hdf..symp..219D} Dickinson, M.\ 1998, The 
Hubble Deep Field, 219
\bibitem[Dickinson(1999)]{1999AIPC..470..122D} Dickinson, M.\ 1999, After 
the Dark Ages: When Galaxies were Young (the Universe at 2 $< Z <$
5), 470, 122
\bibitem[Dickinson et al.(2004)]{2004ApJ...600L..99D} Dickinson, M.~et al.\ 
2004, \apjl, 600, L99
\bibitem[Dickinson \& GOODS Team(2004)]{2004AAS...204.3313D} Dickinson, M., 
\& GOODS Team 2004, Bulletin of the American Astronomical Society, 36, 701 
\bibitem[Dunkley et al.(2008)]{2008arXiv0803.0586D} Dunkley, J., et al.\ 
2008, \apjs, submitted, arXiv:0803.0586 
\bibitem[Efstathiou et al.(1988)]{1988MNRAS.232..431E} Efstathiou, G., 
Ellis, R.~S., \& Peterson, B.~A.\ 1988, \mnras, 232, 431
\bibitem[Erb et al.(2006)]{2006ApJ...646..107E} Erb, D.~K., Steidel, C.~C., 
Shapley, A.~E., Pettini, M., Reddy, N.~A., \& Adelberger, K.~L.\ 2006, 
\apj, 646, 107
\bibitem[Eyles et al.(2007)]{2007MNRAS.374..910E} Eyles, L.~P., Bunker, 
A.~J., Ellis, R.~S., Lacy, M., Stanway, E.~R., Stark, D.~P., \& Chiu, K.\ 
2007, \mnras, 374, 910 
\bibitem[Ford et al.(2003)]{2003SPIE.4854...81F} Ford, H.~C., et al.\ 2003, 
\procspie, 4854, 81 
\bibitem[Gabasch et al.(2004)]{2004A&A...421...41G} Gabasch, A., et al.\ 
2004, \aap, 421, 41
\bibitem[Giavalisco et al.(2004)]{2004ApJ...600L..93G} Giavalisco, M., et 
al.\ 2004a, \apjl, 600, L93
\bibitem[Giavalisco et al.(2004)]{2004ApJ...600L.103G} Giavalisco, M., et 
al.\ 2004b, \apjl, 600, L103
\bibitem[Granato et al.(2004)]{2004ApJ...600..580G} Granato, G.~L., De 
Zotti, G., Silva, L., Bressan, A., \& Danese, L.\ 2004, \apj, 600, 580 
\bibitem[Hathi et al.(2008)]{2008ApJ...673..686H} Hathi, N.~P., Malhotra, 
S., \& Rhoads, J.~E.\ 2008, \apj, 673, 686 
\bibitem[Iwata et al.(2003)]{2003PASJ...55..415I} Iwata, I., Ohta, K., 
Tamura, N., Ando, M., Wada, S., Watanabe, C., Akiyama, M., \& Aoki, K.\ 
2003, \pasj, 55, 415 
\bibitem[Iwata et al.(2007)]{2007MNRAS.376.1557I} Iwata, I., Ohta, K., 
Tamura, N., Akiyama, M., Aoki, K., Ando, M., Kiuchi, G., \& Sawicki, M.\ 
2007, \mnras, 376, 1557 
\bibitem[Iye et al.(2006)]{2006Natur.443..186I} Iye, M., et al.\ 2006, 
\nat, 443, 186 
\bibitem[Kajisawa et al.(2006)]{2006PASJ...58..951K} Kajisawa, M., et al.\ 
2006, \pasj, 58, 951 
\bibitem[Knapp et al.(2004)]{2004AJ....127.3553K} Knapp, G.~R., et al.\ 
2004, \aj, 127, 3553 
\bibitem[Kneib et al.(2004)]{2004ApJ...607..697K} Kneib, J.-P., Ellis, 
R.~S., Santos, M.~R., \& Richard, J.\ 2004, \apj, 607, 697
\bibitem[Kron(1980)]{1980ApJS...43..305K} Kron, R.\ G.\ 1980, \apjs, 43, 
305
\bibitem[labbe et al.(2006)]{2006astro.ph..8444L} Labb\'{e}, I., Bouwens, R., 
Illingworth, G.~D., \& Franx, M.\ 2006, \apjl, 649, 67
\bibitem[Lee et al.(2006)]{2006ApJ...642...63L} Lee, K.-S., Giavalisco, M., 
Gnedin, O.~Y., Somerville, R.~S., Ferguson, H.~C., Dickinson, M., \& Ouchi, 
M.\ 2006, \apj, 642, 63 
\bibitem[Lehnert \& Bremer(2003)]{2003ApJ...593..630L} Lehnert, M.~D.~\& 
Bremer, M.\ 2003, \apj, 593, 630
\bibitem[Madau et al.\ 1998]{mad98} Madau, P., Pozzetti, L. \&
Dickinson, M. 1998, \apj, 498, 106
\bibitem[Magee et al.(2007)]{2007ASPC..376..261M} Magee, D.~K., Bouwens, 
R.~J., \& Illingworth, G.~D.\ 2007, Astronomical Data Analysis Software and Systems XVI, 376, 261 
\bibitem[Mannucci et al. (2007)]{mannucci} Mannucci, F., Buttery, H.,
Maiolino, R., Marconi, A. \& Pozzetti, L. 2007, \aap, 461, 423
\bibitem[Martin et al.(2005)]{2005ApJ...619L..59M} Martin, D.~C., et
al.\ 2005, \apjl, 619, L59
\bibitem[Meurer et al.(1999)]{1999ApJ...521...64M} Meurer, G.~R., Heckman, 
T.~M., \& Calzetti, D.\ 1999, \apj, 521, 64 
\bibitem[Miyazaki et al.(2002)]{2002PASJ...54..833M} Miyazaki, S., et al.\ 
2002, \pasj, 54, 833
\bibitem[Mo \& White(1996)]{1996MNRAS.282..347M} Mo, H.~J., \& White, 
S.~D.~M.\ 1996, \mnras, 282, 347
\bibitem[Mu{\~n}oz \& Loeb(2007)]{2007arXiv0711.2515M} Mu{\~n}oz,
J.~A., \& Loeb, A.\ 2008, \mnras, submitted, arXiv:0711.2515
\bibitem[Oesch et al.(2007)]{2007ApJ...671.1212O} Oesch, P.~A., et al.\ 
2007, \apj, 671, 1212 
\bibitem[Oesch et al.(2008)]{2008arXiv0804.4874O} Oesch, P.~A., et al.\ 
2008, \apj, submitted, arXiv:0804.4874 
\bibitem[Oke \& Gunn(1983)]{1983ApJ...266..713O} Oke, J.~B., \& Gunn, 
J.~E.\ 1983, \apj, 266, 713 
\bibitem[Oppenheimer \& Dav{\'e}(2006)]{2006MNRAS.373.1265O} Oppenheimer, 
B.~D., \& Dav{\'e}, R.\ 2006, \mnras, 373, 1265
\bibitem[Ouchi et al.(2004)]{2004ApJ...611..660O} Ouchi, M., et al.\ 2004a, 
\apj, 611, 660
\bibitem[Ouchi et al.(2004)]{2004ApJ...611..685O} Ouchi, M., et al.\ 2004b, 
\apj, 611, 685
\bibitem[Ouchi et al.(2007)]{2007ASPC..379...47O} Ouchi, M., Tokoku,
C., Shimasaku, K., \& Ichikawa, T.\ 2007, Astronomical Society of the
Pacific Conference Series, 379, 47
\bibitem[Pell{\'o} et al.(2004)]{2004A&A...416L..35P} Pell{\'o}, R., 
Schaerer, D., Richard, J., Le Borgne, J.-F., \& Kneib, J.-P.\ 2004, \aap, 
416, L35
\bibitem[Reddy et al.(2006)]{2006ApJ...644..792R} Reddy, N.~A., Steidel, 
C.~C., Fadda, D., Yan, L., Pettini, M., Shapley, A.~E., Erb, D.~K., \& 
Adelberger, K.~L.\ 2006, \apj, 644, 792 
\bibitem[Reddy et al.(2007)]{2007arXiv0706.4091R} Reddy, N.~A., Steidel, 
C.~C., Pettini, M., Adelberger, K.~L., Shapley, A.~E., Erb, D.~K., \& 
Dickinson, M.\ 2007, arXiv:0706.4091 
\bibitem[Rees \& Ostriker(1977)]{1977MNRAS.179..541R} Rees, M.~J., \& 
Ostriker, J.~P.\ 1977, \mnras, 179, 541 
\bibitem[Richard et al.(2006)]{2006A&A...456..861R} Richard, J., Pell{\'o}, 
R., Schaerer, D., Le Borgne, J.-F., \& Kneib, J.-P.\ 2006, \aap, 456, 861
\bibitem[Riess et al.(2004)]{2004ApJ...607..665R} Riess, A.~G., et al.\ 
2004, \apj, 607, 665 
\bibitem[Riess et al.(2007)]{2007ApJ...659...98R} Riess, A.~G., et al.\ 
2007, \apj, 659, 98
\bibitem[Ryan et al.(2005)]{2005ApJ...631L.159R} Ryan, R.~E., Jr., Hathi, 
N.~P., Cohen, S.~H., \& Windhorst, R.~A.\ 2005, \apjl, 631, L159
\bibitem[Sandage, Tammann, \& Yahil(1979)]{1979ApJ...232..352S} Sandage, 
A., Tammann, G.~A., \& Yahil, A.\ 1979, \apj, 232, 352
\bibitem[Sawicki \& Thompson(2006)]{2006ApJ...642..653S} Sawicki, M., \& 
Thompson, D.\ 2006, \apj, 642, 653
\bibitem[Scannapieco \& Oh(2004)]{2004ApJ...608...62S} Scannapieco, E., \& 
Oh, S.~P.\ 2004, \apj, 608, 62
\bibitem[Schaerer et al.(2007)]{2007IAUS..235..425S} Schaerer, D., Hempel, 
A., Pello, R., Egami, E., Richard, J., Kneib, J.-P., \& Wise, M.\ 2007, IAU 
Symposium, 235, 425 
\bibitem[Schiminovich et al.(2005)]{2005ApJ...619L..47S} Schiminovich, D.,
et al.\ 2005, \apjl, 619, L47 
\bibitem[Sheth \& Tormen(1999)]{1999MNRAS.308..119S} Sheth, R.~K.~\& 
Tormen, G.\ 1999, \mnras, 308, 119
\bibitem[Shimasaku et al.(2005)]{2005PASJ...57..447S} Shimasaku, K., Ouchi, 
M., Furusawa, H., Yoshida, M., Kashikawa, N., \& Okamura, S.\ 2005, \pasj, 
57, 447
\bibitem[Silk(1977)]{1977ApJ...211..638S} Silk, J.\ 1977, \apj, 211, 638 
\bibitem[Skinner et al.(1998)]{1998SPIE.3354....2S} Skinner, C.~J., et al.\ 
1998, \procspie, 3354, 2 
\bibitem[Somerville et al.(2004)]{2004ApJ...600L.171S} Somerville, R.~S., 
Lee, K., Ferguson, H.~C., Gardner, J.~P., Moustakas, L.~A., \& Giavalisco, 
M.\ 2004, \apjl, 600, L171
\bibitem[Spergel et al.(2007)]{2007ApJS..170..377S} Spergel, D.~N., et al.\ 
2007, \apjs, 170, 377 
\bibitem[Springel et al.(2005)]{2005Natur.435..629S} Springel, V., et al.\ 
2005, \nat, 435, 629
\bibitem[Stanway, Bunker, \& McMahon(2003)]{2003MNRAS.342..439S}
Stanway, E.~R., Bunker, A.~J., \& McMahon, R.~G.\ 2003, \mnras, 342,
439
\bibitem[Stanway et al.(2004)]{2004ApJ...607..704S} Stanway, E.~R., Bunker, 
A.~J., McMahon, R.~G., Ellis, R.~S., Treu, T., \& McCarthy, P.~J.\ 2004, 
\apj, 607, 704
\bibitem[Stanway et al.(2005)]{2005MNRAS.359.1184S} Stanway, E.~R., 
McMahon, R.~G., \& Bunker, A.~J.\ 2005, \mnras, 359, 1184
\bibitem[Stanway et al.(2008)]{2008arXiv0801.4559S} Stanway, E.~R., Bremer, 
M.~N., Squitieri, V., Douglas, L.~S., 
\& Lehnert, M.~D.\ 2008, \mnras, in press, arXiv:0801.4559 
\bibitem[Stark et al.(2007)]{2007ApJ...659...84S} Stark, D.~P., Bunker, 
A.~J., Ellis, R.~S., Eyles, L.~P., \& Lacy, M.\ 2007a, \apj, 659, 84 
\bibitem[Stark et al.(2007)]{2007ApJ...663...10S} Stark, D.~P., Ellis, 
R.~S., Richard, J., Kneib, J.-P., Smith, G.~P., \& Santos, M.~R.\ 2007b, 
\apj, 663, 10
\bibitem[Stark et al.(2007)]{2007ApJ...668..627S} Stark, D.~P., Loeb, A., 
\& Ellis, R.~S.\ 2007c, \apj, 668, 627 
\bibitem[Steidel et al.\ (1999)]{1999ApJ...519....1S} Steidel, C.\ C.,
Adelberger, K.\ L., Giavalisco, M., Dickinson, M.\ and Pettini, M.\ 1999,
\apj, 519, 1
\bibitem[Strolger et al.(2004)]{2004ApJ...613..200S} Strolger, L.-G., et 
al.\ 2004, \apj, 613, 200 
\bibitem[Szalay et al.(1999)]{1999AJ....117...68S} Szalay, A.~S.,
Connolly, A.~J., \& Szokoly, G.~P.\ 1999, \aj, 117, 68
\bibitem[Thompson et al.(1998)]{1998ApJ...492L..95T} Thompson, R.~I., 
Rieke, M., Schneider, G., Hines, D.~C., \& Corbin, M.~R.\ 1998, \apjl, 492, 
L95
\bibitem[Thompson et al.(1999)]{1999AJ....117...17T} Thompson, R.~I., 
Storrie-Lombardi, L.~J., Weymann, R.~J., Rieke, M.~J., Schneider, G., 
Stobie, E., \& Lytle, D.\ 1999, \aj, 117, 17 
\bibitem[Thompson et al.(2005)]{2005AJ....130....1T} Thompson, R.~I., et 
al.\ 2005, \aj, 130, 1
\bibitem[Trenti \& Stiavelli(2008)]{2008ApJ...676..767T} Trenti, M.,
\& Stiavelli, M.\ 2008, \apj, 676, 767
\bibitem[Wang \& Heckman(1996)]{1996ApJ...457..645W} Wang, B., \& Heckman, 
T.~M.\ 1996, \apj, 457, 645
\bibitem[Wiklind et al.(2007)]{2007arXiv0710.0406W} Wiklind, T., Dickinson, 
M., Ferguson, H.~C., Giavalisco, M., Mobasher, B., Grogin, N.~A., \& 
Panagia, N.\ 2007, in press, arXiv:0710.0406
\bibitem[Wyder et al.(2005)]{2005ApJ...619L..15W} Wyder, T.~K., et al.\ 
2005, \apjl, 619, L15 
\bibitem[Wyithe \& Loeb(2006)]{2006Natur.441..322W} Wyithe, J.~S.~B., \& 
Loeb, A.\ 2006, \nat, 441, 322
\bibitem[Yan \& Windhorst(2004)]{2004ApJ...612L..93Y} Yan, H.~\& Windhorst, 
R.~A.\ 2004, \apjl, 612, L93
\bibitem[Yan et al.(2005)]{2005ApJ...634..109Y} Yan, H., et al.\ 2005, 
\apj, 634, 109
\bibitem[Yang et al.(2003)]{2003MNRAS.339.1057Y} Yang, X., Mo, H.~J., \& 
van den Bosch, F.~C.\ 2003, \mnras, 339, 1057
\bibitem[Yoshida, M.]{2004} Yoshida, M., et al.\ 2006, \apj, 653, 988
\end{thebibliography}
\end{document}